\batchmode
\makeatletter
\def\input@path{{\string"E:/Trabajo Angel/Mis articulos/Improved LMB approx/Accepted/\string"/}}
\makeatother
\documentclass[british,american]{IEEEtran}
\usepackage[T1]{fontenc}
\usepackage[latin9]{inputenc}
\usepackage{float}
\usepackage{amsthm}
\usepackage{amsmath}
\usepackage{amssymb}
\usepackage{graphicx}
\usepackage{esint}

\makeatletter

\providecommand{\tabularnewline}{\\}
\floatstyle{ruled}
\newfloat{algorithm}{tbp}{loa}
\providecommand{\algorithmname}{Algorithm}
\floatname{algorithm}{\protect\algorithmname}

\theoremstyle{plain}
\newtheorem{thm}{\protect\theoremname}
\theoremstyle{definition}
\newtheorem{defn}[thm]{\protect\definitionname}
\theoremstyle{plain}
\newtheorem{prop}[thm]{\protect\propositionname}

\usepackage{cite} 
\usepackage[margin=8pt,font=footnotesize]{caption}
\usepackage{algorithm}
\usepackage{algpseudocode}
\allowdisplaybreaks 
\@ifundefined{showcaptionsetup}{}{%
 \PassOptionsToPackage{caption=false}{subfig}}
\usepackage{subfig}
\makeatother

\usepackage{babel}
\addto\captionsamerican{\renewcommand{\definitionname}{Definition}}
\addto\captionsamerican{\renewcommand{\propositionname}{Proposition}}
\addto\captionsamerican{\renewcommand{\theoremname}{Theorem}}
\addto\captionsbritish{\renewcommand{\algorithmname}{Algorithm}}
\addto\captionsbritish{\renewcommand{\definitionname}{Definition}}
\addto\captionsbritish{\renewcommand{\propositionname}{Proposition}}
\addto\captionsbritish{\renewcommand{\theoremname}{Theorem}}
\providecommand{\definitionname}{Definition}
\providecommand{\propositionname}{Proposition}
\providecommand{\theoremname}{Theorem}

\begin{document}

\title{A track-before-detect labelled multi-Bernoulli particle filter with
label switching}

\author{Ángel F. García-Fernández\thanks{Copyright (c) 2016 IEEE. Personal use of this material is permitted. Permission from IEEE must be obtained for all other users, including reprinting/ republishing this material for advertising or promotional purposes, creating new collective works for resale or redistribution to servers or lists, or reuse of any copyrighted components of this work in other works.

The author is with the Department of Electrical and Computer Engineering, Curtin University, Perth, WA 6102, Australia (email: angel.garciafernandez@curtin.edu.au). 

This work was supported in part by the Australian Research Council under Discovery Project DP130104404.}}
\maketitle
\begin{abstract}
This paper presents a multitarget tracking particle filter (PF) for
general track-before-detect measurement models. The PF is presented
in the random finite set framework and uses a labelled multi-Bernoulli
approximation. We also present a label switching improvement algorithm
based on Markov chain Monte Carlo that is expected to increase filter
performance if targets get in close proximity for a sufficiently long
time. The PF is tested in two challenging numerical examples.

\end{abstract}

\begin{IEEEkeywords}
Particle filters, MCMC, random finite sets, multitarget tracking,
Kullback-Leibler divergence
\end{IEEEkeywords}

\section{Introduction}

In surveillance applications it is important to accurately estimate
the number of targets along with their states at each time step based
on a sequence of measurements. A relevant difficulty in this multitarget
tracking (MTT) problem is the fact that the number of targets is unknown
and time varying. Yet, Bayesian inference on multiobject systems,
such as MTT, can be done in a mathematically rigorous way using the
random finite set (RFS) formulation \cite{Mahler_book14}. Here, the
multitarget state is a set that contains the single target states
and all the available information about the targets is included in
the posterior probability density function (PDF), i.e., the PDF of
the state given all available measurements \cite{Mahler03}. In most
cases of interest, calculating the posterior PDF is intractable due
to nonlinear/non-Gaussian dynamic and measurement models and the difficulty
of handling target births and deaths. Consequently, it must be approximated.

If a unique identifying label is added to each single target state,
we can estimate the states of specific targets at different time steps
\cite{Vo13,Angel13}. Moreover, the resulting Bayesian recursion is
more easily performed because of the simplification of the set integrals
in the prediction step \cite{Vo13,Angel15_e}. This implies that it
can be useful to include target labels, which could be considered
as auxiliary variables, even if we are only interested in performing
inference on the unlabelled collection of targets. Importantly in
this case, we have an extra degree of freedom that can be used to
improve the posterior PDF approximation: we can choose any labelled
posterior PDF as long as the corresponding unlabelled posterior PDF
remains unaltered. This kind of idea, which we refer to as label switching
improvement, was first introduced for fixed and known number of targets
and Gaussian approximations in \cite{Svensson11}. It can lead to
benefits in performance if there is mixed labelling \cite{Boers10},
i.e., when targets get in close proximity for a sufficiently long
time and then separate. An extension of \cite{Svensson11} for MTT,
with Gaussian multi-Bernoulli PDFs, has been proposed in \cite{Williams15}. 

In this paper, we first design an efficient particle filter (PF) \cite{Ristic_book04}
for track-before-detect MTT with no assumptions in the measurement
model. Then, we develop an algorithm for label switching improvement
for this PF based on Markov chain Monte Carlo (MCMC) \cite{Liu_book01}.
We proceed to review the literature and explain our contributions
more thoroughly. 

As we focus on general track-before-detect applications, MTT algorithms
for the radar point detection measurement model, such as the probability
hypothesis density (PHD) filter \cite{Mahler03}, cardinalised PHD
(CPHD) filter \cite{Mahler07}, multiple hypothesis tracking \cite{Blackman_book99},
the labelled RFS filters in \cite{Vo13,Reuter14,Vo14} or the sequential
Monte Carlo (SMC) algorithms in \cite{Vermaak05,Sarkka07} cannot
be applied. There are multiple track-before-detect MTT algorithms
that are not general as they require specific measurement models such
as superpositional sensors \cite{Nannuru13,Papi15}, models with likelihood
factorisation over single targets \cite{Vo10}, pixelised sensors
\cite{Kreucher05,Morelande07,Fallon12} or the model of the histogram
probabilistic multi-hypothesis tracker \cite{Davey13}.

We address the track-before-detect problem for general measurement
models by considering an approximation to the posterior PDF based
on SMC methods or particle filters (PFs). Markov chain Monte Carlo
(MCMC) methods can also be applied to perform Bayesian filtering with
the same flexibility as PFs \cite{Septier09} but we focus on PFs.
A PF provides an SMC approximation to the posterior which converges
to the posterior as the number of particles tends to infinity under
some conditions \cite{Del_Moral_inbook00}. In practice however, it
is desirable to keep the number of particles low to reduce the computational
burden as well as increasing the speed of the filter. Due to the curse
of dimensionality, direct generalisations of single target PFs to
MTT do not work well for a reasonably low number of particles \cite{Angel13}.
One way of alleviating this problem is to make the posterior independence
assumption (PIA), in which target states are independent \cite{Orton02,Kreucher05,Angel13,Yi13,Ubeda14}.
While PIA implies that the PF is no longer asymptotically optimal,
it is usually beneficial for low sample sizes \cite{Yi13}. 

In order to account for the unknown and variable number of targets
in PFs, one possibility is to sample target existences directly from
the prior \cite{Boers04}. Yet, this is highly inefficient as it removes
and adds targets regardless of the current measurement. In \cite{Kreucher05,Morelande07,Fallon12},
the current measurement is taken into account to draw samples from
target existences by an existence grid, and in \cite{Angel13}, by
a two-layer PF. However, in these cases, sampling of target existences
and states is done via separate procedures, which can imply detrimental
effects on performance if several targets are in close proximity.
In this paper, we handle target states and label existences jointly
using the labelled RFS framework by assuming a labelled multi-Bernoulli
(LMB) posterior, which implies that target states and existences are
independent \cite{Vo13}. The proposed PF, the generalised parallel
partition (GPP) PF, is an extension of the parallel partition (PP)
PF, which was developed for fixed and known number of targets \cite[Sec. III]{Angel13}.
In contrast to the PP method, the GPP method does not require a two-layer
PF \cite[Sec. IV]{Angel13} to take into account variable target number
as the target states and existences are handled jointly. We generalise
the PP method due to its computational efficiency and high performance
\cite{Angel13,Yi13,Ubeda14}. 

As mentioned before, we can make use of label switching improvement
techniques, which are useful if targets move in close proximity for
a long time and we are not interested in labelling information \cite{Angel14}.
In this respect, we propose an algorithm to obtain a PDF that does
not alter the unlabelled target information and can be more accurately
approximated as LMB than the original posterior PDF. This improvement
in the LMB approximation enhances the performance of the particle
filter. This algorithm is based on iterated Kullback-Leibler divergence
(KLD) minimisations and is a generalisation of \cite{Angel14_b},
which considers a fixed and known number of targets. 

The remainder of the paper is organised as follows. We formulate the
problem in Section \ref{sec:Problem-formulation}. Section \ref{sec:Improvement-LMB}
explains the recursion for improving the LMB approximation. In Section
\ref{sec:GPP filter}, we develop the GPP particle filter. The implementation
of the label switching improvement algorithm for the particle filter
is developed in Section \ref{sec:MCMC-algorithm}. Numerical examples
are provided in Section \ref{sec:Numerical-simulations}. Finally,
conclusions are drawn in Section \ref{sec:Conclusions}.

\section{Problem formulation\label{sec:Problem-formulation}}

In this paper, labelled RFS densities are denoted as $\boldsymbol{\pi}$,
unlabelled RFS densities as $\check{\pi}$ and densities over a vector
space as $\pi$, which are referred to as vector densities. A brief
introduction to the RFS framework with unlabelled and labelled sets
can be found in Sections II and III in \cite{Vo13}. Without loss
of generality, we remove the time index of the filtering recursion
for notational simplicity. Variables and densities at the previous
time step have a superscript $^{-}$. 

The collection of targets at the current time step is represented
by the unlabelled set $X=\left\{ x_{1},...,x_{t}\right\} $, where
$x_{i}\in\mathbb{R}^{n_{x}}$ is the state of the $i$th target. The
multiobject transition density $\check{f}\left(\cdot\left|X^{-}\right.\right)$
encapsulates the underlying models of target dynamics, births and
deaths. Targets are observed through noisy measurements, which can
be vectors or sets. Once the measurement has been observed, the resulting
multi-target likelihood $\ell\left(\cdot\right)$ depends on $X$
and with this notation we highlight that it is not a PDF on $X$.
For the sake of notational simplicity, we omit the explicit value
of the measurement in the likelihood.

The objective of this paper is to compute the unlabelled RFS posterior
density $\check{\pi}$, whose argument is $X$, for general track-before-detect
measurement models. The PDF $\check{\pi}$ contains all information
regarding the unlabelled states given the sequence of measurements.
The filtering recursion is \cite{Mahler_book14}
\begin{align}
\check{\pi}\left(X\right) & \propto\ell\left(X\right)\int\check{f}\left(X\left|X^{-}\right.\right)\check{\pi}^{-}\left(X^{-}\right)\delta X^{-}\label{eq:filtering_set}
\end{align}
where $\propto$ denotes proportionality.

Even though our objective is to calculate the unlabelled RFS posterior,
using a labelled posterior is beneficial to perform the filtering
recursion because set integrals are easier to compute \cite{Vo13,Angel15_e}.
In this case, labels can be seen as auxiliary variables that aid in
computation. We denote the labelled set as $\mathbf{X}=\left\{ \left(x_{1},l_{1}\right),...,\left(x_{t},l_{t}\right)\right\} $
where $l_{i}\in\mathbb{L}$ is the label for the $i$th target, no
two targets can have the same label and $\mathbb{L}$ is a countable
set. The filtering recursion becomes \cite{Mahler_book14}
\begin{align}
\boldsymbol{\pi}\left(\mathbf{X}\right) & \propto\ell\left(\mathbf{X}\right)\int\boldsymbol{f}\left(\mathbf{X}\left|\mathbf{X}^{-}\right.\right)\boldsymbol{\pi}^{-}\left(\mathbf{X}^{-}\right)\delta\mathbf{X}^{-}\label{eq:filtering_label}
\end{align}
where $\boldsymbol{\pi}$ is the labelled RFS posterior at the current
time step and $\boldsymbol{f}\left(\cdot\left|\mathbf{X}^{-}\right.\right)$
is the labelled RFS transition density. It should be highlighted that
$\boldsymbol{f}\left(\cdot\left|\mathbf{X}^{-}\right.\right)$ has
the property that the labels of the surviving targets do not change
with time. This is the main reason why the set integrals in (\ref{eq:filtering_label})
are easier to compute than in (\ref{eq:filtering_set}). Based on
\cite{Svensson11}, we make the following definition.
\begin{defn}
\label{def:The-unlabelled-RFS-family}The unlabelled RFS family $\left[\boldsymbol{\pi}\right]$
of $\boldsymbol{\pi}$ is $\left[\boldsymbol{\pi}\right]=\left\{ \boldsymbol{\varphi}:\boldsymbol{\varphi}\sim\boldsymbol{\pi}\right\} $,
where $\boldsymbol{\varphi}\sim\boldsymbol{\pi}$ if and only if $\boldsymbol{\varphi}$
and $\boldsymbol{\pi}$ have the same unlabelled PDF, which is obtained
by integrating out the labels \cite[Eq. (9)]{Vo13}:
\begin{align*}
 & \sum_{\left(l_{1},...,l_{t}\right)\in\mathbb{L}^{t}}\boldsymbol{\varphi}\left(\left\{ \left(x_{1},l_{1}\right),...,\left(x_{t},l_{t}\right)\right\} \right)\\
 & \quad=\sum_{\left(l_{1},...,l_{t}\right)\in\mathbb{L}^{t}}\boldsymbol{\pi}\left(\left\{ \left(x_{1},l_{1}\right),...,\left(x_{t},l_{t}\right)\right\} \right)\:\forall t\in\mathbb{N}.
\end{align*}

\end{defn}
Set $\left[\boldsymbol{\pi}\right]$ is an equivalence class as $\sim$
is reflexive, symmetric and transitive. If the targets move independently
with the same dynamics and measurements do not provide labelling information,
we can calculate (\ref{eq:filtering_set}) using (\ref{eq:filtering_label})
by a labelled RFS density $\boldsymbol{\pi}^{-}$ such that if we
integrate out the labels we obtain $\check{\pi}^{-}$. Importantly,
at any time step, we can change the labelled RFS density $\boldsymbol{\pi}^{-}$
with another labelled density $\boldsymbol{\varphi}^{-}\in\left[\boldsymbol{\pi}^{-}\right]$
at our convenience. These ideas were first used in \cite{Svensson11}
to provide more accurate Gaussian approximations in MTT with fixed
and known number of targets. 

\begin{figure}
\begin{centering}
\includegraphics[scale=0.7]{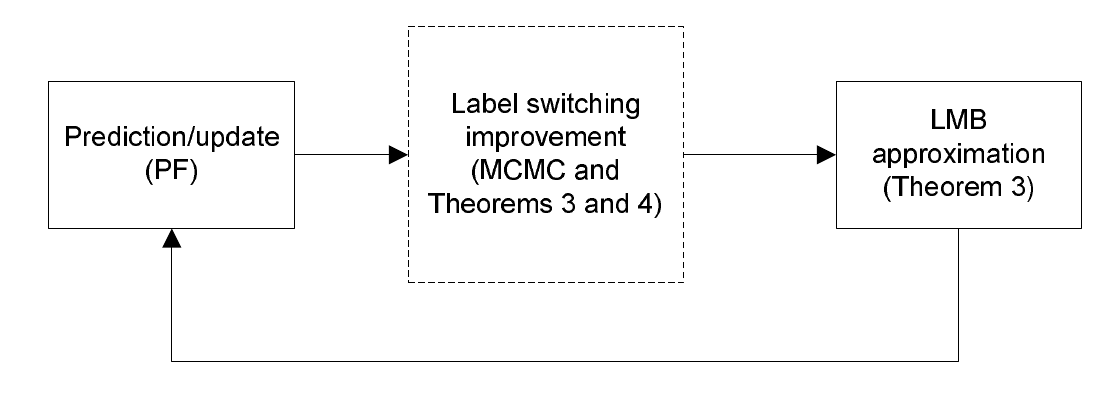}
\par\end{centering}

\protect\caption{\label{fig:Diagram}Diagram of the proposed LMB-PF recursion. The
label switching improvement algorithm is optional as it is only useful
if targets get in close proximity for a sufficiently long time and
then separate.}
\end{figure}

In order to approximate the unlabelled posterior for general measurement
models, this paper proposes a PF with LMB approximation. The prediction
and update step are therefore performed via Monte Carlo sampling and
the resulting PDF is approximated as LMB via KLD minimisation. Instead
of performing the LMB approximation on the labelled posterior provided
by the PF, we can choose another labelled PDF within its unlabelled
RFS family that can be approximated as LMB more accurately to improve
performance. This algorithm is designed based on KLD minimisations
and implemented via MCMC. The resulting LMB-PF recursion is illustrated
in Figure \ref{fig:Diagram}. The label switching improvement algorithm
is explained in Section \ref{sec:Improvement-LMB}, the PF in Section
\ref{sec:GPP filter} and the implementation of the label switching
improvement for the PF based on MCMC in Section \ref{sec:MCMC-algorithm}. 

\section{Improvement of LMB approximation\label{sec:Improvement-LMB}}

In this section we describe the label switching improvement algorithm,
see Figure \ref{fig:Diagram}. That is, given a labelled RFS density
$\boldsymbol{\pi}$, we provide a recursion of PDFs to select a labelled
PDF that belongs to $\left[\boldsymbol{\pi}\right]$ that is more
accurately approximated as LMB than $\boldsymbol{\pi}$. First, we
write a simplified expression for the KLD for labelled RFS in Section
\ref{sub:Kullback-Leibler-divergence}. Second, we review LMB RFS
in Section \ref{sub:Labelled-multi-Bernoulli}. Third, we provide
the required iterated optimisations in Section \ref{sub:Iterated-optimisations}.
Fourth, in Section \ref{sub:Illustrative-example}, we provide an
example to illustrate the results of this section.

\subsection{Kullback-Leibler divergence\label{sub:Kullback-Leibler-divergence}}

In this section, we derive a decomposition of the KLD for labelled
sets. First, we introduce the required notation. Given a labelled
RFS density $\boldsymbol{\pi}$ and $t$ different labels $l_{1},...,l_{t}$,
we define the PDF (over a vector space) \cite{Angel15_e}
\begin{align}
\pi\left(x_{1:t};l_{1},...,l_{t}\right) & \triangleq\frac{\boldsymbol{\pi}\left(\left\{ \left(x_{1},l_{1}\right),...,\left(x_{t},l_{t}\right)\right\} \right)}{P_{\pi}\left(\left\{ l_{1},...,l_{t}\right\} \right)}\label{eq:pdf_conditioned_states}
\end{align}
where
\begin{align}
P_{\pi}\left(\left\{ l_{1},...,l_{t}\right\} \right) & =\int\boldsymbol{\pi}\left(\left\{ \left(x_{1},l_{1}\right),...,\left(x_{t},l_{t}\right)\right\} \right)dx_{1:t}\label{eq:pdf_labels}
\end{align}
is the probability of having a labelled set with labels $\left\{ l_{1},...,l_{t}\right\} $
and $x_{1:t}=\left(x_{1},...,x_{t}\right)$. 
\begin{prop}
\label{prop:KLD}Given two labelled RFS densities $\boldsymbol{\pi}$
and $\boldsymbol{\nu}$, the KLD \cite{Mahler_book14}
\begin{align}
\mathrm{D}\left(\boldsymbol{\pi}\left\Vert \boldsymbol{\nu}\right.\right)= & \int\boldsymbol{\pi}\left(\mathbf{X}\right)\log\frac{\boldsymbol{\pi}\left(\mathbf{X}\right)}{\boldsymbol{\nu}\left(\mathbf{X}\right)}\delta\mathbf{X}
\end{align}
can be written as
\begin{align}
\mathrm{D}\left(\boldsymbol{\pi}\left\Vert \boldsymbol{\nu}\right.\right) & =\mathrm{D}\left(P_{\pi}\left\Vert P_{\nu}\right.\right)\nonumber \\
 & \quad+\sum_{L\subseteq\mathbb{L}}P_{\pi}\left(L\right)\mathrm{D}\left(\pi\left(\cdot;\overrightarrow{L}\right)\left\Vert \nu\left(\cdot;\overrightarrow{L}\right)\right.\right)\label{eq:KLD}
\end{align}
where $\overrightarrow{L}$ is a vector whose components are the elements
of $L$ arranged in ascending order.
\end{prop}
Proposition \ref{prop:KLD} is proved in Appendix \ref{sec:AppendixA}.
We want to remark that Proposition \ref{prop:KLD} holds for any ordering
$\overrightarrow{L}$ of $L$. As pointed out in Section \ref{sub:Iterated-optimisations},
we use labels that are natural numbers, as in \cite{Angel13,Angel09},
so we can use ascending order without loss of generality. If labels
are vectors as in \cite{Vo13}, we can define $\overrightarrow{L}$
based on  lexicographical order \cite{Kolman_book01}.

\subsection{Labelled multi-Bernoulli RFS\label{sub:Labelled-multi-Bernoulli}}

An LMB density $\boldsymbol{\nu}$ is defined under the assumption 
\begin{itemize}
\item A1 There is a maximum number $\kappa$ of targets.
\end{itemize}
An LMB RFS includes a target with label $j\in\mathbb{L}$ with probability
$p_{j}$ and, if this target exists, its state is distributed according
to a density $\nu\left(\cdot;j\right)$ independently from the rest
of targets \cite{Vo13}. In this paper, we characterise its RFS density
by
\begin{align}
P_{\nu}\left(\left\{ l_{1},...,l_{t}\right\} \right) & =\prod_{j=1}^{\kappa}\left(1-p_{j}\right)\prod_{j=1}^{t}\frac{p_{l_{j}}}{1-p_{l_{j}}}\label{eq:constraint_ind_existences}\\
\nu\left(x_{1:t};l_{1:t}\right) & =\nu\left(x_{1};l_{1}\right)...\nu\left(x_{t};l_{t}\right)\label{eq:constraint_ind_PDFs}
\end{align}
where we have assumed without loss of generality that $\mathbb{L}=\left\{ 1,...,\kappa\right\} $.

\subsection{Iterated optimisations\label{sub:Iterated-optimisations}}

In this section, ideally, we would like to find the best LMB approximation
$\boldsymbol{\nu}$ to any density $\boldsymbol{\varphi}\in\left[\boldsymbol{\pi}\right]$.
Therefore, given $\boldsymbol{\pi}$, we would like to find
\begin{align}
\boldsymbol{\nu}^{\star} & =\underset{\boldsymbol{\nu}}{\arg\min}\,\mathrm{D}\left(\boldsymbol{\varphi}\left\Vert \boldsymbol{\nu}\right.\right)\label{eq:optimisation_problem}
\end{align}
subject to $\boldsymbol{\varphi}\in\left[\boldsymbol{\pi}\right]$
and $\boldsymbol{\nu}$ is LMB. However, this optimisation problem
is difficult to solve.

Instead, in this paper, we provide a sequence of PDFs $\boldsymbol{\varphi}^{0}=\boldsymbol{\pi}$,
$\boldsymbol{\nu}^{0}$, $\boldsymbol{\varphi}^{1}$, $\boldsymbol{\nu}^{1}$,...
such that at each step of the iteration the KLD is lowered. Specifically,
based on $\boldsymbol{\varphi}^{n}$, $\boldsymbol{\nu}^{n}$ is obtained
by minimising $\mathrm{D}\left(\boldsymbol{\varphi}^{n}\left\Vert \boldsymbol{\nu}^{n}\right.\right)$
with constraints (\ref{eq:constraint_ind_existences}) and (\ref{eq:constraint_ind_PDFs}).
How to perform this minimisation will be indicated by Theorem \ref{thm:optimisation_independent}.
Note that the result of this theorem provides us with the best LMB
approximation to a labelled RFS density according to the KLD. Given
$\boldsymbol{\nu}^{n}$, $\boldsymbol{\varphi}^{n+1}$ is calculated
as follows. First, the algorithm sets $P_{\varphi^{n+1}}=P_{\pi}$
so that the original probability mass function (PMF) of the labels
does not change. Second, we set $\varphi^{n+1}\left(\cdot;l{}_{1:t}\right)$=$\varphi^{n}\left(\cdot;l_{1:t}\right)$
for all the labels $\left\{ l_{1},...,l_{t}\right\} \subseteq\left\{ 1,...,\kappa\right\} $.
Then, we go through all the labels and modify $\varphi^{n+1}\left(\cdot;l_{1:t}\right)$
by minimising $\mathrm{D}\left(\boldsymbol{\varphi}^{n+1}\left\Vert \boldsymbol{\nu}^{n}\right.\right)$
with constraint $\boldsymbol{\varphi}\in\left[\boldsymbol{\pi}\right]$.
How to perform these minimisations will be indicated by Theorem \ref{thm:Optimisation_RFS_family}.
The resulting sequence is quite suitable for PF implementations, as
will be seen in Section \ref{sec:MCMC-algorithm}. Due to how the
minimisations are performed, 
\begin{equation}
\mathrm{D}\left(\boldsymbol{\varphi}^{n}\left\Vert \boldsymbol{\nu}^{n}\right.\right)\geq\mathrm{D}\left(\boldsymbol{\varphi}^{n+1}\left\Vert \boldsymbol{\nu}^{n}\right.\right)\geq\mathrm{D}\left(\boldsymbol{\varphi}^{n+1}\left\Vert \boldsymbol{\nu}^{n+1}\right.\right).
\end{equation}
This implies that the final LMB approximation is equal or more accurate
than the original. As $\mathrm{D}\left(\boldsymbol{\varphi}\left\Vert \boldsymbol{\nu}\right.\right)\geq0$,
this sequence converges. The required optimisations are given below. 
\begin{thm}
\label{thm:optimisation_independent}For a given labelled RFS density
$\boldsymbol{\varphi}$, the solution to
\[
\underset{\boldsymbol{\nu}}{\arg\min}\,\mathrm{D}\left(\boldsymbol{\varphi}\left\Vert \boldsymbol{\nu}\right.\right)
\]
subject to $\boldsymbol{\nu}$ being LMB, which implies (\ref{eq:constraint_ind_existences})
and (\ref{eq:constraint_ind_PDFs}), is 
\begin{align}
p_{j} & =\sum_{L\ni j}P_{\varphi}\left(L\right)\\
\nu\left(x_{1};j\right) & \propto\sum_{L\ni j}P_{\varphi}\left(L\right)\varphi_{j}\left(x_{1};\overrightarrow{L}\right)\quad j=\left\{ 1,...,\kappa\right\} 
\end{align}
where $\varphi_{j}\left(\cdot;\overrightarrow{L}\right)$ denotes
the marginal PDF of the state that corresponds to label $j$ given
the labels $\overrightarrow{L}$. 
\end{thm}
This theorem is proved in Appendix \ref{sec:AppendixB}. We want to
remark that $\varphi\left(\cdot;\overrightarrow{L}\right)$ is a vector
density so the marginal $\varphi_{j}\left(\cdot;\overrightarrow{L}\right)$
simply corresponds to integrating out the states except the one with
label $j$. In \cite[Sec. III.B]{Reuter14}, an LMB approximation
from a $\delta$-generalised LMB PDF is proposed. Theorem 3 implies
that the approximation provided in \cite{Reuter14} actually minimises
the KLD in that particular case. 
\begin{thm}
\label{thm:Optimisation_RFS_family}For given labelled RFS densities
$\boldsymbol{\nu}$ and $\boldsymbol{\pi}$, PMF $P_{\varphi}=P_{\pi}$,
label set $L$ and vector densities $\varphi\left(\cdot;\overrightarrow{L'}\right)$
$L'\subseteq\mathbb{L}\setminus L$ the solution to
\[
\underset{\varphi\left(\cdot;\overrightarrow{L}\right)}{\arg\min}\,\mathrm{D}\left(\boldsymbol{\varphi}\left\Vert \boldsymbol{\nu}\right.\right)
\]
subject to $\boldsymbol{\varphi}\in\left[\boldsymbol{\pi}\right]$
is 
\begin{align}
\varphi\left(x_{1:\left|L\right|};\overrightarrow{L}\right) & =\alpha\left(x_{1:\left|L\right|};\overrightarrow{L}\right)\check{\pi}\left(\left\{ x_{1},...,x_{\left|L\right|}\right\} ;\overrightarrow{L}\right)\\
\alpha\left(x_{1:\left|L\right|};\overrightarrow{L}\right) & =\frac{\nu\left(x_{1:\left|L\right|};\overrightarrow{L}\right)}{\check{\nu}\left(\left\{ x_{1},...x_{\left|L\right|}\right\} ;\overrightarrow{L}\right)}
\end{align}
where the RFS density given the labels is
\begin{align}
\check{\pi}\left(\left\{ x_{1},...x_{\left|L\right|}\right\} ;\overrightarrow{L}\right) & =\sum_{p=1}^{\left|L\right|!}\pi\left(\Gamma_{p,\left|L\right|}\left(x_{1:\left|L\right|}\right);\overrightarrow{L}\right)\label{eq:RFS_density_conditioned_labels}
\end{align}
and $\Gamma_{p,t}\left(\cdot\right)$ indicates the $p$th permutation
for $t$ elements.
\end{thm}
Theorem \ref{thm:Optimisation_RFS_family} is proved in Appendix \ref{sec:AppendixC}.
We want to clarify that, according to (\ref{eq:pdf_conditioned_states})-(\ref{eq:pdf_labels}),
$\boldsymbol{\varphi}$ is characterised by $P_{\varphi}$ and $\varphi\left(\cdot;\overrightarrow{L}\right)$
$L\subseteq\mathbb{L}$. In Theorem \ref{thm:Optimisation_RFS_family},
we are given all these densities except one, which is the one we optimise.
Finally, the steps of the recursive optimisations are given in Algorithm
\ref{alg:LMB_improvement}.

\begin{algorithm}
\protect\caption{\label{alg:LMB_improvement}Improved LMB approximation }

{\fontsize{9}{9}\selectfont

\textbf{Input:} Initial labelled density $\boldsymbol{\pi}$.

\textbf{Output:} LMB approximation $\boldsymbol{\nu}^{\star}$.

\begin{algorithmic}     

\State - Set $\boldsymbol{\varphi}^{0}=\boldsymbol{\pi}$.

\For{ $n=0$ to $I_{1}-1$ } \Comment{ $I_{1}$ \textit{is the
number of steps.} } 

\State - Calculate $\boldsymbol{\nu}^{n}$ using $\boldsymbol{\varphi}^{n}$
and Theorem \ref{thm:optimisation_independent}. 

\State - Set $\boldsymbol{\varphi}^{n+1}=\boldsymbol{\varphi}^{n}$.

\ForAll{$L\subseteq\mathbb{L}$ }

\State - Calculate $\varphi^{n+1}\left(\cdot;\overrightarrow{L}\right)$
using $\boldsymbol{\nu}^{n}$, $\varphi^{n+1}\left(\cdot;\overrightarrow{L'}\right)$ 

\State \quad{}$L'\neq L$ and Theorem \ref{thm:Optimisation_RFS_family}. 

\EndFor

\EndFor

\State - Set $\boldsymbol{\nu}^{\star}=\boldsymbol{\nu}^{I_{1}-1}$.

\end{algorithmic}

}
\end{algorithm}

\subsection{Illustrative example\label{sub:Illustrative-example}}

In this section we consider an illustrative example to show how the
previous algorithm for lowering the KLD works. Let us assume $\kappa=2$,
\begin{align}
\pi\left(x_{1:2};1,2\right) & =\mathcal{N}\left(x_{1:2};\left[10,11\right]^{T},\left[\begin{array}{cc}
\sigma_{1}^{2} & \rho\sigma_{1}\sigma_{2}\\
\rho\sigma_{1}\sigma_{2} & \sigma_{2}^{2}
\end{array}\right]\right)\\
\pi\left(x_{1};1\right) & =\mathcal{N}\left(x_{1};10,\sigma_{1}^{2}\right)\\
\pi\left(x_{1};2\right) & =\mathcal{N}\left(x_{1};11,\sigma_{2}^{2}\right)
\end{align}
where $\sigma_{1}=\sigma_{2}=1$, $\rho=-0.8$, and $\mathcal{N}\left(x;\overline{x},\Sigma\right)$
is the Gaussian PDF with mean $\overline{x}$ and covariance matrix
$\Sigma$ evaluated at $x$. We consider two cases that differ in
the PMF of the labels, see Table \ref{tab:PMF-illustrative}. An important
feature is that $P_{\pi}\left(\left\{ 1,2\right\} \right)$ is considerably
larger than the rest in Case 1.

\begin{table}
\protect\caption{\label{tab:PMF-illustrative}PMF $P_{\pi}$ for the illustrative example}

\centering{}%
\begin{tabular}{c|cccc}
\hline 
 &
$\textrm{Ø}$ &
$\left\{ 1\right\} $ &
$\left\{ 2\right\} $ &
$\left\{ 1,2\right\} $\tabularnewline
\hline 
Case 1 &
0.1 &
0.05 &
0.05 &
0.8\tabularnewline
Case 2 &
0.1 &
0.3 &
0.3 &
0.3\tabularnewline
\hline 
\end{tabular}
\end{table}

\begin{figure}
\begin{raggedright}
\subfloat[]{\protect\begin{centering}
\protect\includegraphics[scale=0.3]{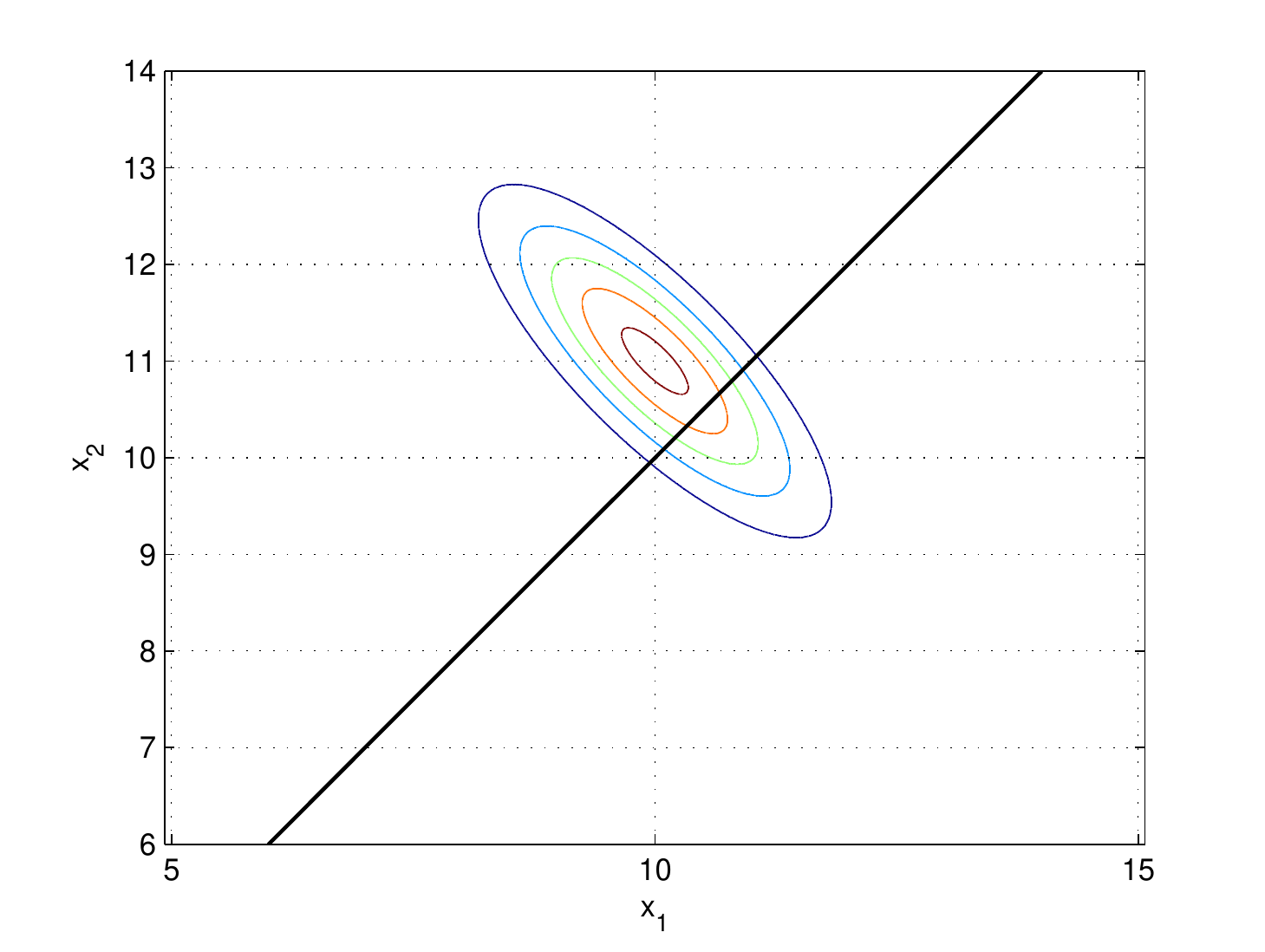}\protect
\par\end{centering}

}\subfloat[]{\protect\begin{centering}
\protect\includegraphics[scale=0.3]{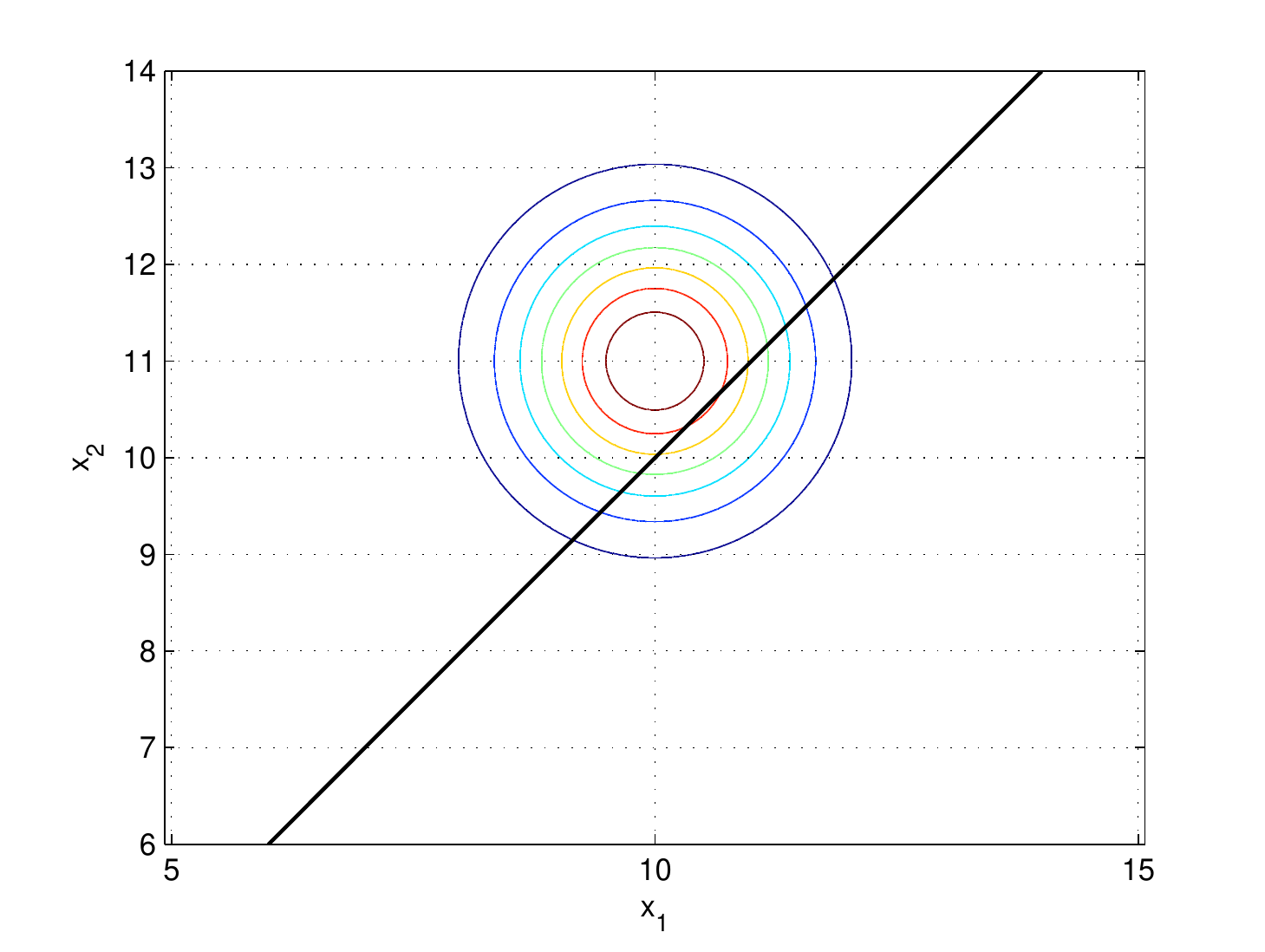}\protect
\par\end{centering}

}
\par\end{raggedright}

\protect\caption{\label{fig:PDFs_ini}Joint PDF for labels $1,2$ for Cases 1 and 2:
(a) initial PDF $\varphi^{0}\left(\cdot;1,2\right)$ and (b) its best
LMB approximation $\nu^{0}\left(\cdot;1,2\right)$}

\end{figure}

\begin{figure}
\begin{raggedright}
\subfloat[]{\protect\begin{centering}
\protect\includegraphics[scale=0.3]{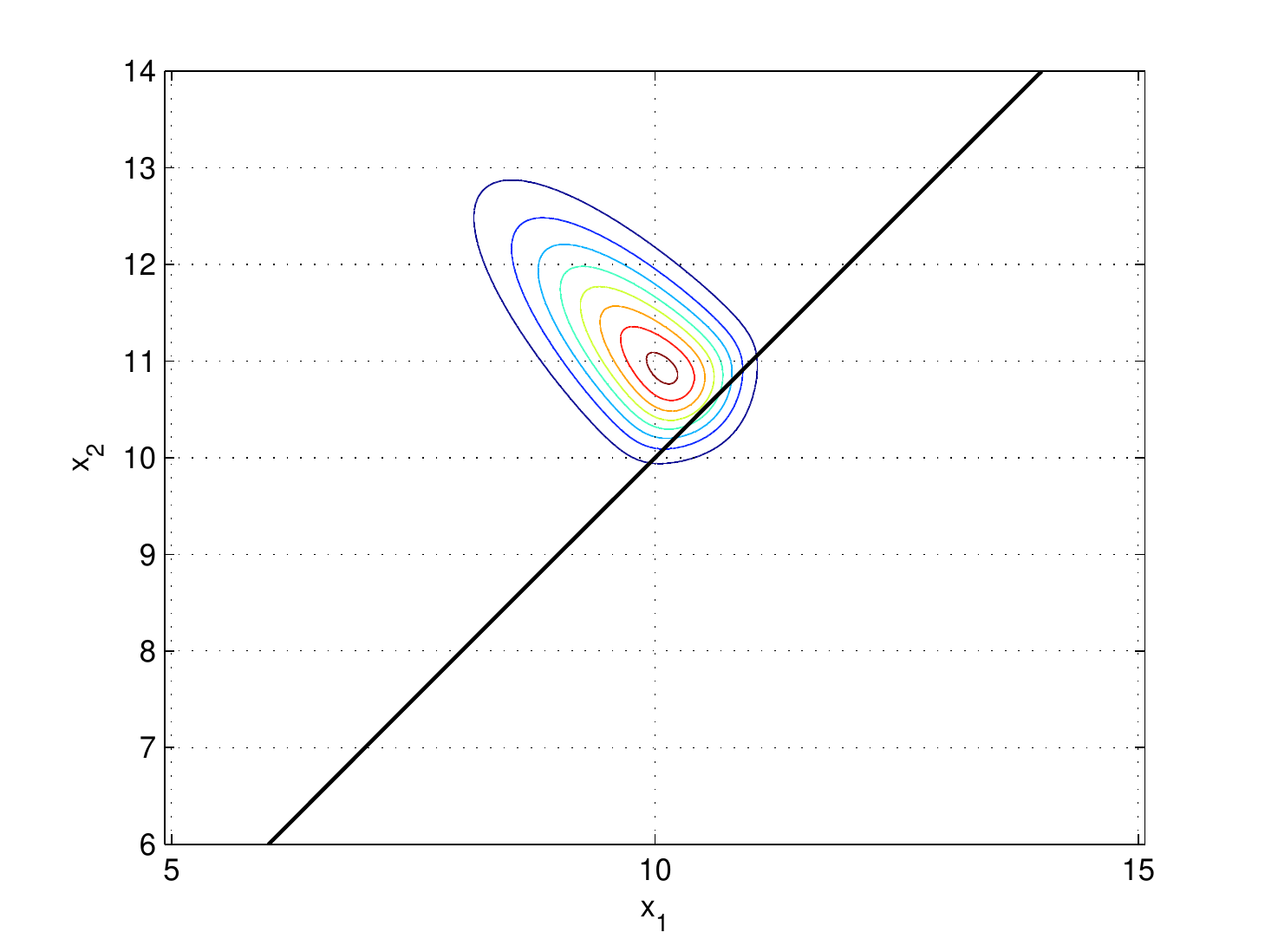}\protect
\par\end{centering}

}\subfloat[]{\protect\begin{centering}
\protect\includegraphics[scale=0.3]{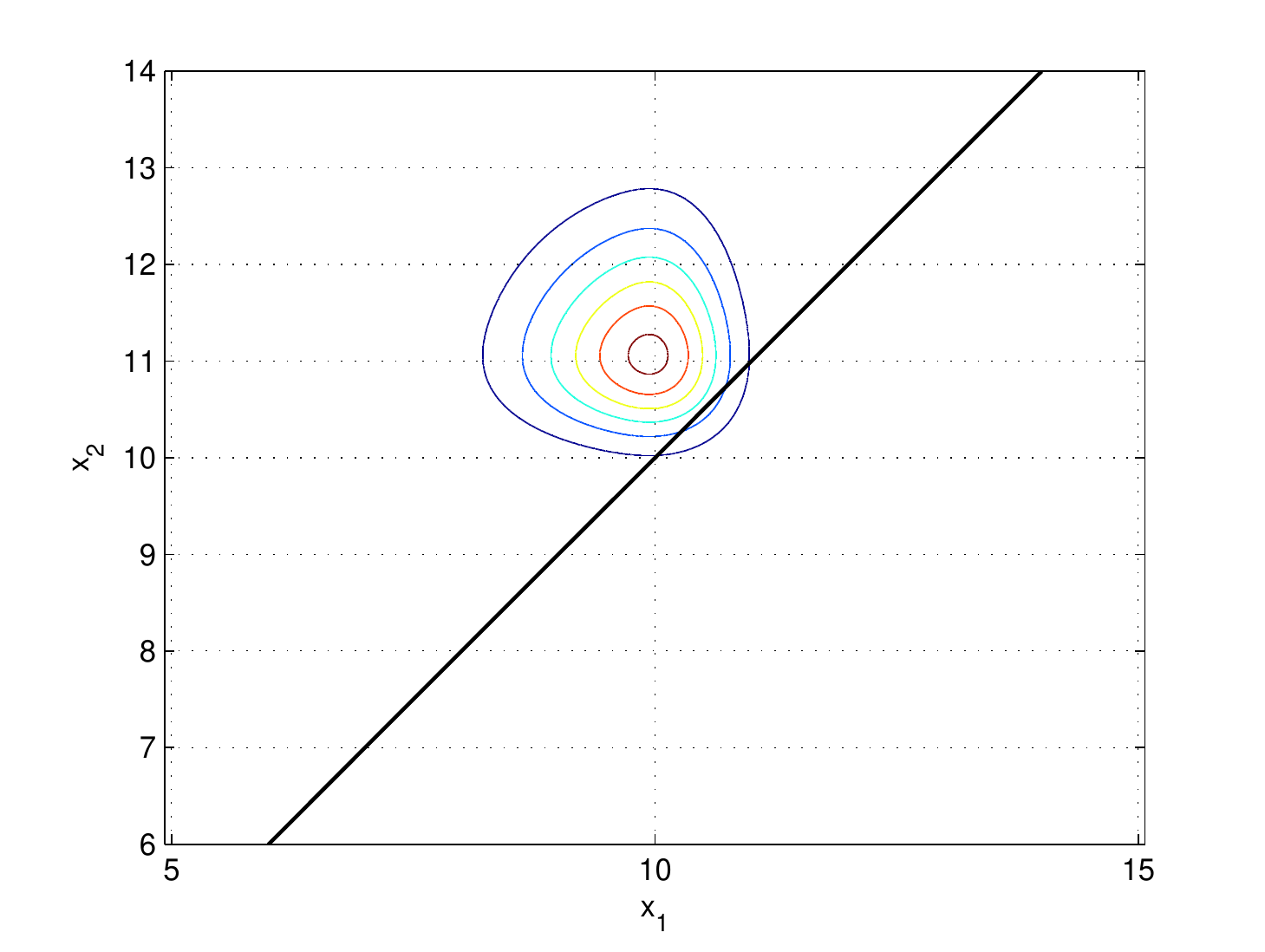}\protect
\par\end{centering}

}
\par\end{raggedright}

\protect\caption{\label{fig:Final_PDFs_case1}Joint PDF for labels $1,2$ for Case
1: (a) PDF $\varphi^{5}\left(\cdot;1,2\right)$ and (b) its best LMB
approximation $\nu^{5}\left(\cdot;1,2\right)$}
\end{figure}

The PDFs for labels 1 and 2 for the original PDF $\boldsymbol{\varphi}^{0}$
and its best LMB approximation $\boldsymbol{\nu}^{0}$ are shown in
Figure \ref{fig:PDFs_ini}. Both cases have the same PDF for labels
1 and 2 in the best LMB approximation because $\pi\left(\cdot;1\right)$,
$\pi\left(\cdot;2\right)$ correspond with the marginal PDFs of $\pi\left(\cdot;1,2\right)$,
although this is not necessarily this case. It should be noted that
in both cases, $\varphi^{n}\left(\cdot;1\right)=\pi\left(\cdot;1\right)$
and $\varphi^{n}\left(\cdot;2\right)=\pi\left(\cdot;2\right)$ $\forall n$.
This is due to the fact that the previous recursion only performs
changes in the PDFs that represent more than one target. As we iterate,
the KLD gets lower as shown in Table \ref{tab:Kullback-Leibler-divergence}.
The PDFs for labels 1 and 2 for $\boldsymbol{\varphi}^{5}$ and $\boldsymbol{\nu}^{5}$
in Case 1 are shown in Figure \ref{fig:Final_PDFs_case1}. It is clear
that there are considerable differences between $\boldsymbol{\varphi}^{0}$
and $\boldsymbol{\varphi}^{5}$ that enable the significant lowering
of the KLD, although they contain the same information regarding the
corresponding unlabelled set. In case 2, the KLD is also reduced but
much less, see Table \ref{tab:Kullback-Leibler-divergence}.  The
reason behind this behaviour is that, in Case 1, it is highly likely
that two targets exist and the weight of the PDFs with labels with
a single target are quite low. Therefore, the resulting optimisation
is quite similar to the case in which there are always two targets,
see \cite{Angel14_b}. On the contrary, in Case 2, the single target
PDFs have a more important weight and the best LMB approximation is
clearly influenced by $\pi\left(\cdot;1\right)$ and $\pi\left(\cdot;2\right)$
so the KLD cannot be reduced much. Nevertheless, the recursion always
ensures that the new PDFs are more accurately approximated as LMB
than the original ones. 

\begin{table}
\protect\caption{\label{tab:Kullback-Leibler-divergence}KLD of $\boldsymbol{\nu}^{n}$
from $\boldsymbol{\varphi}^{n}$}

\centering{}%
\begin{tabular}{c|cccccc}
\hline 
Iteration number $n$ &
0 &
1 &
2 &
3 &
4 &
5\tabularnewline
\hline 
Case 1 &
0.542 &
0.467 &
0.408 &
0.382 &
0.375 &
0.374\tabularnewline
Case 2 &
0.184 &
0.164 &
0.160 &
0.159 &
0.159 &
0.159\tabularnewline
\hline 
\end{tabular}
\end{table}

\section{Labelled multi-Bernoulli particle filter\label{sec:GPP filter}}

In this section, we describe the generalised parallel partition (GPP)
particle filter, which is a generalisation of the parallel partition
(PP) particle filter \cite[Sec. III]{Angel13} that accounts for target
births and deaths. The PP-PF corresponds to the GPP-PF with a known
and fixed number of targets. Note that for fixed and known number
of targets, a vector representation of the multitarget state is equivalent
to a labelled RFS \cite{Angel14}, so, in this case, an LMB RFS density
is equivalent to a density on a multitarget vector state with independent
targets. In Section \ref{sub:Posterior-density}, we provide the posterior
PDF and explain the use of an auxiliary variable in GPP-PF. The importance
density to sample the posterior and the resulting particle weights
are provided in Section \ref{sub:Importance-density}.

\subsection{Posterior density\label{sub:Posterior-density}}

We assume
\begin{itemize}
\item A2 The posterior at the previous time step is LMB with existence probabilities
$p_{j}^{-}$ and densities $\pi^{-}\left(\cdot;j\right)$ $j\in\left\{ 1,...,\kappa^{-}\right\} $.
\item A3 The PDF of the new born targets is LMB with existence probabilities
$p_{j}$ and densities $\eta\left(\cdot;j\right)$ $j\in\left\{ \kappa^{-}+1,...,\kappa\right\} $.
\end{itemize}
We recall that we assume without loss of generality that $\mathbb{L}=\left\{ 1,...,\kappa\right\} $.
In a PF, we draw $N$ particles $\left\{ \mathbf{X}^{1},...,\mathbf{X}^{N}\right\} $
from an importance density, where $\mathbf{X}^{i}$ represents the
$i$th particle, and in order to evaluate the weight of the $i$th
particle we need to evaluate $\boldsymbol{\pi}\left(\mathbf{X}^{i}\right)$.
Note that these particles are usually stored in matrices in a computer
implementation \cite[Sec. III.A]{Angel15_e}. In the following, we
represent the posterior PDF using decomposition (\ref{eq:pdf_conditioned_states})-(\ref{eq:pdf_labels}).
For evaluating the posterior for a specific particle, we find it useful
to denote by $\overrightarrow{L^{i}}$ and $x_{1:\left|L^{i}\right|}^{i}$
the vector of labels of $\mathbf{X}^{i}$ arranged in ascending order
and $x_{1:\left|L^{i}\right|}^{i}$ their corresponding states \cite{Angel15_e}.
Therefore,
\begin{align}
\boldsymbol{\pi}\left(\mathbf{X}^{i}\right) & =P_{\pi}\left(L^{i}\right)\pi\left(x_{1:\left|L^{i}\right|}^{i};\overrightarrow{L^{i}}\right).
\end{align}

Under A2 and considering Section \ref{sub:Labelled-multi-Bernoulli},
the posterior at the previous time is characterised by
\begin{align}
P_{\pi^{-}}\left(L\right) & =\prod_{j=1}^{\kappa^{-}}\left(1-p_{j}^{-}\right)\prod_{j=1}^{\left|L\right|}\frac{p_{L_{j}}^{-}}{1-p_{L_{j}}^{-}}\\
\pi^{-}\left(x_{1:\left|L\right|};\overrightarrow{L}\right) & =\prod_{j=1}^{\left|L\right|}\pi^{-}\left(x_{j};L_{j}\right)
\end{align}
where $L_{j}$ indicates the $j$th component of $\overrightarrow{L}=\left(L_{1},...,L_{\left|L\right|}\right)$. 

Let us assume that we have a Monte Carlo (MC) approximation of the
posterior at the previous time such that
\begin{align}
\pi^{-}\left(x_{1:\left|L\right|};\overrightarrow{L}\right) & \propto\prod_{j=1}^{\left|L\right|}\sum_{i=1}^{N_{L_{j}}^{-}}\delta\left(x_{j}-x_{L_{j}}^{-i}\right)
\end{align}
where $\delta\left(\cdot\right)$ is the Dirac delta, $x_{L_{j}}^{-i}$
is the $i$th particle and $N_{L_{j}}^{-}$ is the number of particles
of density $\pi^{-}\left(\cdot;L_{j}\right)$ and the weights are
even. Then, making the usual assumptions that targets move independently
with single target transition density $g\left(\cdot|\cdot\right)$
and probability of survival $\gamma$, the prior PDF $\boldsymbol{\xi}$
of the surviving targets, which is also called the predicted PDF of
the surviving targets, is also LMB with \cite{Reuter14}
\begin{align}
P_{\xi}\left(L\right) & =\prod_{j=1}^{\kappa^{-}}\left(1-p_{j}\right)\prod_{j=1}^{\left|L\right|}\frac{p_{L_{j}}}{1-p_{L_{j}}}\\
\xi\left(x_{1:\left|L\right|};\overrightarrow{L}\right) & \propto\prod_{j=1}^{\left|L\right|}\sum_{i=1}^{N_{L_{j}}^{-}}g_{L_{j}}^{i}\left(x_{j}\right)
\end{align}
where
\begin{align}
p_{L_{j}} & =\gamma\cdot p_{L_{j}}^{-}\\
g_{j}^{i}\left(\cdot\right) & =g\left(\cdot|x_{j}^{-i}\right)\quad j\in\left\{ 1,...,\kappa^{-}\right\} .
\end{align}

Under A3, the prior PDF $\boldsymbol{\omega}$ at the current time
step, which is the predicted PDF of the surviving targets and the
new born targets, is also LMB with \cite{Reuter14} 
\begin{align}
P_{\omega}\left(L\right) & =\prod_{j=1}^{\kappa}\left(1-p_{j}\right)\prod_{j=1}^{\left|L\right|}\frac{p_{L_{j}}}{1-p_{L_{j}}}\\
\omega\left(x_{1:\left|L\right|};\overrightarrow{L}\right) & \propto\prod_{j=1}^{\left|L\right|}\sum_{i=1}^{N_{L_{j}}^{-}}g_{L_{j}}^{i}\left(x_{j}\right)\label{eq:prior_existence_particle}
\end{align}
where
\begin{align}
g_{j}^{i}\left(\cdot\right)= & \eta\left(\cdot;j\right)\quad j\in\left\{ \kappa^{-}+1,...,\kappa\right\} \\
N_{L_{j}}^{-}= & N\quad j\in\left\{ \kappa^{-}+1,...,\kappa\right\} .
\end{align}
Note that in (\ref{eq:prior_existence_particle}) we write the PDFs
of the new born targets also as a mixture of PDFs to deal with surviving
and new born targets jointly with the same notation. Applying Bayes'
rule, we obtain the posterior \cite{Angel15_e}
\begin{align}
\pi\left(x_{1:\left|L\right|};\overrightarrow{L}\right)P_{\pi}\left(L\right) & \propto\ell\left(\left\{ x_{1},...,x_{\left|L\right|}\right\} \right)\nonumber \\
 & \quad\times P_{\omega}\left(L\right)\prod_{j=1}^{\left|L\right|}\sum_{i=1}^{N_{L_{j}}^{-}}g_{L_{j}}^{i}\left(x_{j}\right).\label{eq:posterior_existence}
\end{align}

As in the PP method \cite{Angel13}, we use an auxiliary vector $a_{1:\left|L\right|}=\left(a_{1},...,a_{\left|L\right|}\right)$
such that we write
\begin{align}
\pi\left(x_{1:\left|L\right|},a_{1:\left|L\right|};\overrightarrow{L}\right)P_{\pi}\left(L\right) & \propto\ell\left(\left\{ x_{1},...,x_{\left|L\right|}\right\} \right)\nonumber \\
 & \quad\times P_{\omega}\left(L\right)\prod_{j=1}^{\left|L\right|}g_{L_{j}}^{a_{j}}\left(x_{j}\right)\label{eq:posterior_PP}
\end{align}
where $a_{j}\in\left\{ 1,...,N_{L_{j}}^{-}\right\} $ and each component
of $a_{1:\left|L\right|}$ is an index on the mixture in (\ref{eq:posterior_existence}).
In other words, in (\ref{eq:posterior_PP}), the state $x_{j}$ of
target $j$ comes from particle $a_{j}$ at the previous time step.
The auxiliary vector is quite useful because it lowers the computational
complexity by removing the sum in (\ref{eq:posterior_existence})
and allows for subparticle crossover \cite{Angel13}. As illustrated
in \cite[Fig. 1]{Angel13}, subparticle crossover refers to the fact
that a (multitarget) particle of the posterior can be formed by propagating
subparticles, part of the multitarget particle that represents a target,
that belonged to different particles at the previous time step. As
expected, integrating out $a_{1:\left|L\right|}$ in (\ref{eq:posterior_PP}),
we obtain (\ref{eq:posterior_existence}). In general, samples from
(\ref{eq:posterior_PP}) cannot be obtained directly, so we proceed
to describe an importance density to draw samples from.

\subsection{Importance density and particle weights\label{sub:Importance-density}}

\begin{figure}
\begin{centering}
\includegraphics[scale=0.5]{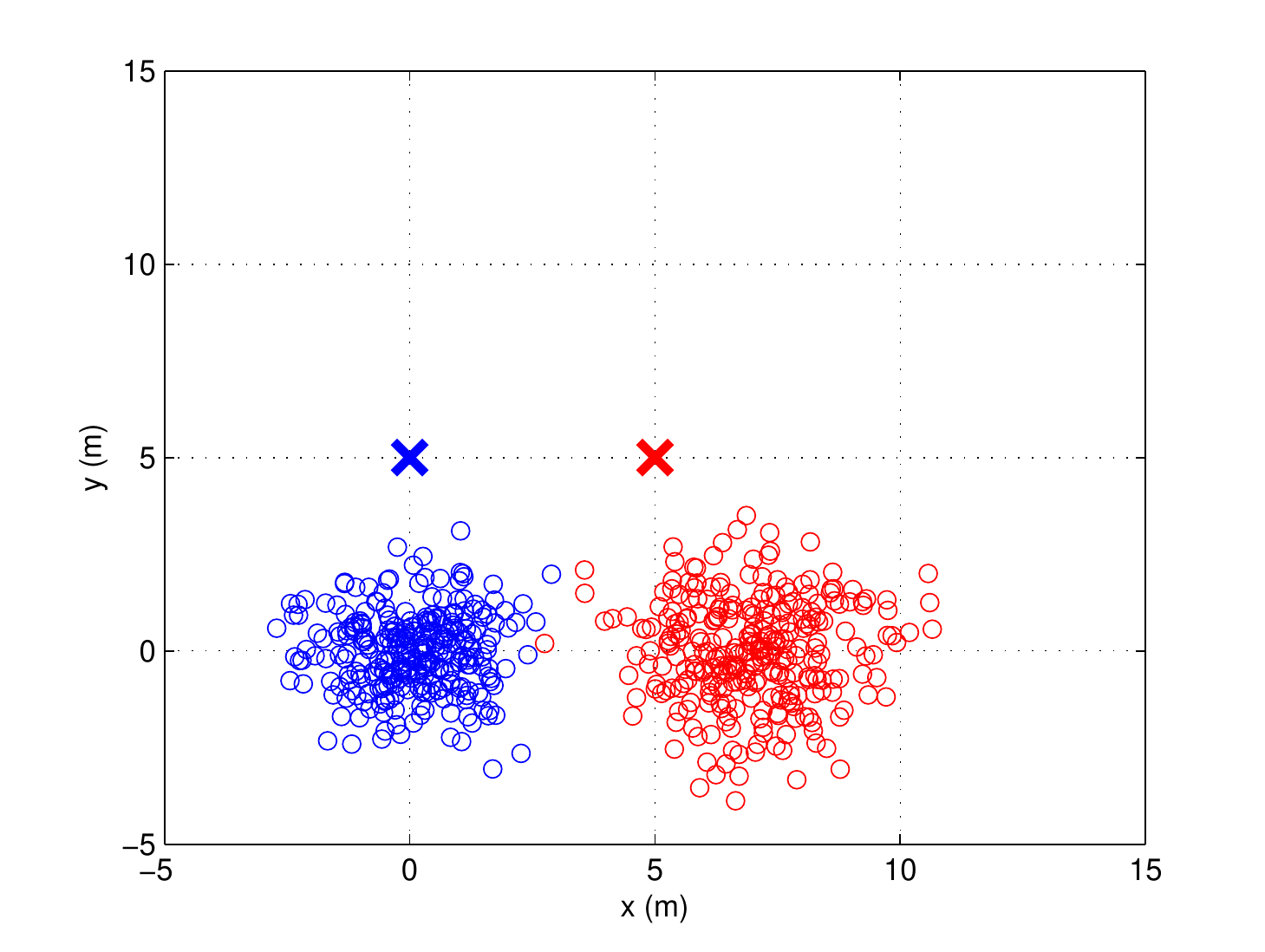}
\par\end{centering}

\protect\caption{\label{fig:Illustration-PP_existence}Illustration of the sampling
of the label set. The position particles of two targets at the previous
time step are shown as blue and red circles. The crosses denote the
predicted positions at the current time step for each target, which
are obtained using these particles and the dynamic model, see (\ref{eq:predicted_state}).
Based on these predicted positions and the current measurement, we
sample the label PMF using (\ref{eq:q_existence}).}
\end{figure}

The predicted state of the target with label $j$ is\footnote{If we cannot obtain $\mathrm{E}\left[g_{j}^{i}\left(x\right)\right]$
in closed-form, we can instead draw $\hat{x}_{j}$ as a sample from
$g_{j}^{i}\left(\cdot\right)$ as discussed in \cite{Ubeda14} for
the PP method or in \cite{Pitt99} for the auxiliary PF.}
\begin{equation}
\hat{x}_{j}=\frac{1}{N_{j}^{k-1}}\sum_{i=1}^{N_{j}^{-}}\mathrm{E}\left[g_{j}^{i}\left(x\right)\right]\quad j\in\left\{ 1,...,\kappa\right\} .\label{eq:predicted_state}
\end{equation}
The set of predicted states with labels in $L$ is
\begin{align*}
\hat{X}_{L} & =\left\{ \hat{x}_{L_{1}},...,\hat{x}_{L_{\left|L\right|}}\right\} .
\end{align*}
We first write the importance density $\boldsymbol{q}$ in terms of
its decomposition (\ref{eq:pdf_conditioned_states})-(\ref{eq:pdf_labels})
and then we explain it:
\begin{align}
P_{q}\left(L\right) & \propto P_{\omega}\left(L\right)\ell\left(\hat{X}_{L}\right)\label{eq:q_existence}\\
q\left(x_{1:\left|L\right|},a_{1:\left|L\right|};\overrightarrow{L}\right) & =\prod_{j=1}^{\left|L\right|}\frac{g_{L_{j}}^{a_{j}}\left(x_{j}\right)\ell\left(\left\{ x_{j}\right\} \cup\hat{X}_{L\setminus L_{j}}\right)}{\beta_{L_{j}}}\label{eq:q_state}\\
\beta_{L_{j}}= & \sum_{a_{j}=1}^{N_{L_{j}}^{-}}\int g_{L_{j}}^{a_{j}}\left(x_{j}\right)\ell\left(\left\{ x_{j}\right\} \cup\hat{X}_{L\setminus L_{j}}\right)dx_{j}\label{eq:gamma_PP}
\end{align}
where $L\setminus L_{j}$ indicates label set $L$ without label $L_{j}$
and $\beta_{L_{j}}$ is a normalising constant. Sampling from is $\boldsymbol{q}$
performed in two steps. First, we obtain $N$ samples $L^{1},...,L^{N}$
from (\ref{eq:q_existence}). As $L$ is a discrete set, this task
can be performed easily. Each of these samples contains the labels
of the targets whose states have to be sampled to obtain the final
particles from the posterior. An important characteristic of $P_{q}$
is how it takes into account the current measurement and target states.
It uses the predicted target states for each configuration of labels
(hypotheses), represented by $\hat{X}_{L}$, and the current measurement,
included in $\ell\left(\hat{X}_{L}\right)$, to draw possible existences,
see Figure \ref{fig:Illustration-PP_existence}. This way, we draw
more particles with hypotheses which are predicted to have a higher
posterior weight. 

Once we have obtained $N$ samples $L^{1},...,L^{N}$ from (\ref{eq:q_existence}),
we have $H$ different label sets $\dot{L}^{1},...,\dot{L}^{H}$ and
$N_{\dot{L}^{h}}$ particles with label set $\dot{L}^{h}$. Then,
for all the particles with label set $\dot{L}^{h}$, we use (\ref{eq:q_state})
to obtain $N_{\dot{L}^{h}}$ samples of the target states. The density
(\ref{eq:q_state}) is the importance density of the PP method given
the label set $\dot{L}^{h}$. How target state sampling is performed
and the properties of the PP method are thoroughly discussed in \cite[Sec. III]{Angel13}.
Target states are sampled independently, this is why there is a product
over the target index $j$, and the other target states are taken
into account in the likelihood by their predicted states, see (\ref{eq:predicted_state})
and Figure \ref{fig:Illustration-PP_existence}. It should be noted
that in (\ref{eq:q_state}), we include the normalising constant $\beta_{L_{j}}$.
The reason behind this is that the importance density (\ref{eq:q_state})
for each $\dot{L}^{h}$ is different so we have to consider the normalising
constants in the final particle weight. The normalising constants
are approximated as \cite{Vermaak05}
\begin{equation}
\beta_{L_{j}}\approx\frac{1}{N_{L_{j}}^{-}}\sum_{i=1}^{N_{L_{j}}^{-}}\ell\left(\left\{ \tilde{x}_{j}^{i}\right\} \cup\hat{X}_{L\setminus L_{j}}\right)\label{eq:normalising_constant}
\end{equation}
where $\tilde{x}_{j}^{i}\sim g_{L_{j}}^{i}\left(\cdot\right)$. 

The final particles are represented by $\left\{ \left(L^{1},x_{1:\left|L^{1}\right|}^{1},a_{1:\left|L^{1}\right|}^{1}\right),....,\left(L^{N},x_{1:\left|L^{N}\right|}^{N},a_{1:\left|L^{N}\right|}^{N}\right)\right\} $.
Using (\ref{eq:posterior_PP}) and (\ref{eq:q_existence})-(\ref{eq:gamma_PP}),
the weight of the $i$th particle becomes
\begin{align}
w^{i} & \propto\frac{\ell\left(\left\{ x_{1}^{i},...,x_{\left|L^{i}\right|}^{i}\right\} \right)}{\ell\left(\hat{X}_{L^{i}}\right)\prod_{j=1}^{\left|L^{i}\right|}b_{j}\left(L^{i},x_{j}^{i}\right)}\label{eq:final_weight_PP}\\
b_{j}\left(L^{i},x_{j}^{i}\right) & =\frac{\ell\left(\left\{ x_{j}^{i}\right\} \cup\hat{X}_{L^{i}\setminus L_{j}^{i}}\right)}{\beta_{L_{j}^{i}}}.\label{eq:weight_b_PP}
\end{align}
where $L_{j}^{i}$ denotes the $j$th component in ascending order
of $L^{i}$, i.e., $\overrightarrow{L}^{i}=\left(L_{1}^{i},...,L_{\left|L^{i}\right|}^{i}\right)$.

The steps for drawing samples in the GPP method are provided in Algorithm
\ref{alg:GPP_method_subroutine}. Finally, in order to be able to
perform the filtering recursion, at the next time step, we need a
prior of the form (\ref{eq:prior_existence_particle}). This can be
achieved by first performing resampling to obtain evenly distributed
weights and directly approximate the current particle approximation
as LMB using Theorem \ref{thm:optimisation_independent}. This results
in an LMB approximation such that the particles and existence probability
for the target with label $j$ are:
\begin{align}
\left\{ x_{j}^{1},...,x_{j}^{N_{j}}\right\}  & =\left\{ x_{p}^{i}:L_{p}^{i}=j,\,p\in\left\{ 1,2,...,\left|L^{i}\right|\right\} \right\} \label{eq:theorem2_application_lmb}\\
p_{j} & =\frac{N_{j}}{N}\label{eq:theorem2_application_existence}
\end{align}
where (\ref{eq:theorem2_application_lmb}) simply takes the particles
which have label $j$ and (\ref{eq:theorem2_application_existence})
indicates that the probability of existence of the $j$target is proportional
to the number of particles that contain it. 

If we are not interested in target labels instead of using (\ref{eq:theorem2_application_lmb})-(\ref{eq:theorem2_application_existence}),
we can use the label switching improvement algorithm explained in
Section \ref{sec:MCMC-algorithm}. The overall steps of the PF are
indicated in Algorithm \ref{alg:Particle-filter-algorithm}.

\begin{algorithm}
\protect\caption{\label{alg:GPP_method_subroutine}GPP particle filter subroutine}

{\fontsize{9}{9}\selectfont

\textbf{Input:} Existence probability\textbf{ }$p_{j}^{-}$ and $N_{j}^{-}$
particles $\left\{ x_{j}^{-1},...,x_{j}^{-N_{j}^{-}}\right\} $ for
$j=\left\{ 1,...,\kappa^{-}\right\} $, which represent (\ref{eq:prior_existence_particle}). 

\textbf{Output:} $N$ particles with even weights $\left\{ \left(L^{1},x_{1:\left|L^{1}\right|}^{1},a_{1:\left|L^{1}\right|}^{1}\right),....,\left(L^{N},x_{1:\left|L^{N}\right|}^{N},a_{1:\left|L^{N}\right|}^{N}\right)\right\} $,
which approximate (\ref{eq:posterior_PP}).

\begin{algorithmic}     

\State - Calculate $p_{j}=\gamma\cdot p_{j}^{-}$ for $j=\left\{ 1,...,\kappa^{-}\right\} $.

\State - Calculate $\hat{x}_{j}$ for $j=\left\{ 1,...,\kappa\right\} $
using (\ref{eq:predicted_state}). \Comment{ \textit{This step includes
new born targets}} 

\State - Obtain $N$ samples $L^{1},...,L^{N}$ from (\ref{eq:q_existence}).

\State - From $L^{1},...,L^{N}$, obtain the different label sets
$\dot{L}^{1},...,\dot{L}^{H}$ and the number $N_{\dot{L}^{h}}$ of
particles with $\dot{L}^{h}$.

\State - Set $r=1$ \Comment{ $r$\textit{ is an index on the particles
of the posterior.} } 

\Comment{ \textit{Go through all the sampled label sets.} } 

\For{ $h=1$ to $H$ }

\Comment{ \textit{Go through the targets indicated by} $\dot{L}^{h}$.
} 

\For{ $j=1$ to $\left|\dot{L}^{h}\right|$ }

\Comment{ \textit{Draw the particles for the $j$th target in} $\dot{L}^{h}$.
} 

\For{ $l=1$ to $N_{\dot{L}_{j}^{h}}^{-}$ }

\State - Obtain a sample $\tilde{x}_{j}^{l}\sim g_{\dot{L}_{j}^{h}}^{l}\left(x_{j}\right)$. 

\State - Calculate $c\left(\tilde{x}_{j}^{l}\right)=\ell\left(\left\{ \tilde{x}_{j}^{l}\right\} \cup\hat{X}_{\dot{L}^{h}\setminus\dot{L}_{j}^{h}}\right)$.

\EndFor

\State - Normalise $c\left(\tilde{x}_{j}^{l}\right)$ to sum to one
over $l=1$ to $N_{\dot{L}_{j}^{h}}^{-}$. 

\For{ $l=1$ to $N_{\dot{L}^{h}}$ }

\Comment{ \textit{Resampling stage for the $j$th target in }$\dot{L}^{h}$.
} 

\State - Sample an index $p$ from the distribution defined by 

\State \qquad{}$c\left(\tilde{x}_{j}^{i}\right)$ over $i=1$ to
$N_{\dot{L}_{j}^{h}}^{-}$.

\State - Set $x_{j}^{r}=\tilde{x}_{j}^{p}$, $L_{j}^{r}=\dot{L}_{j}^{h}$
and $a_{j}^{r}=p$.

\State - Set $b_{j}\left(\dot{L}_{j}^{h},x_{j}^{r}\right)=N_{\dot{L}_{j}^{h}}^{-}\cdot c\left(\tilde{x}_{j}^{p}\right)$. 

\State - Set $r=r+1$.

\EndFor

\EndFor

\EndFor

\Comment{ \textit{Weight calculation} } 

\For{ $i=1$ to $N$ }

\State - Calculate the weight $w^{i}$ using (\ref{eq:final_weight_PP}).

\EndFor

\State - Perform resampling to obtain even particle weights.

\end{algorithmic}

}
\end{algorithm}

\begin{algorithm}

\protect\caption{\label{alg:Particle-filter-algorithm}LMB particle filter algorithm}

{\fontsize{9}{9}\selectfont

\textbf{Input:} Existence probability\textbf{ }$p_{j}^{-}$ and $N_{j}^{-}$
particles $\left\{ x_{j}^{-1},...,x_{j}^{-N_{j}^{-}}\right\} $ for
$j=\left\{ 1,...,\kappa^{-}\right\} $, which represent (\ref{eq:prior_existence_particle}). 

\textbf{Output:} Existence probability\textbf{ }$p_{j}$ and $N_{j}$
particles $\left\{ x_{j}^{1},...,x_{j}^{N_{j}}\right\} $ for $j=\left\{ 1,...,\kappa\right\} $,
which approximate (\ref{eq:posterior_PP}) as LMB.

\begin{algorithmic}     

\State - Use Algorithm \ref{alg:GPP_method_subroutine} to obtain
$N$ particles with even weights $\left\{ \left(L^{1},x_{1:\left|L^{1}\right|}^{1},a_{1:\left|L^{1}\right|}^{1}\right),....,\left(L^{N},x_{1:\left|L^{N}\right|}^{N},a_{1:\left|L^{N}\right|}^{N}\right)\right\} $,
which approximate (\ref{eq:posterior_PP}).

\Comment{ \textit{Obtain LMB approximation} } 

\If{Label switching improvement}

\Comment{ \textit{Useful when targets get in close proximity} } 

\State - Use Algorithm \ref{alg:ILMB-MCMC_algorithm} to obtain $p_{j}$
and $N_{j}$ particles $\left\{ x_{j}^{1},...,x_{j}^{N_{j}}\right\} $
for $j=\left\{ 1,...,\kappa\right\} $.

\Else

\State - Use (\ref{eq:theorem2_application_lmb})-(\ref{eq:theorem2_application_existence})
to obtain $p_{j}$ and $N_{j}$ particles $\left\{ x_{j}^{1},...,x_{j}^{N_{j}}\right\} $
for $j=\left\{ 1,...,\kappa\right\} $.

\EndIf

\end{algorithmic}

}

\end{algorithm}

\section{Label switching improvement using MCMC\label{sec:MCMC-algorithm}}

In this section, we apply the label switching algorithm to improve
the LMB approximation described in Section \ref{sec:Improvement-LMB}
to the output of the PF developed in the previous section, see Figure
\ref{fig:Diagram}. More specifically, we use MCMC to approximate
the sequence $\boldsymbol{\varphi}^{0}=\boldsymbol{\pi}$, $\boldsymbol{\nu}^{0}$,
$\boldsymbol{\varphi}^{1}$, $\boldsymbol{\nu}^{1}$... indicated
in Algorithm \ref{alg:LMB_improvement}. The resulting algorithm is
called improved LMB MCMC (ILMB-MCMC). 

First, we calculate the RFS density given the labels, which is given
by (\ref{eq:RFS_density_conditioned_labels}), using the posterior
(\ref{eq:posterior_existence}) to obtain 
\begin{align}
\check{\pi}\left(\left\{ x_{1},...,x_{\left|L\right|}\right\} ;\overrightarrow{L}\right) & \propto\ell\left(\left\{ x_{1},...,x_{\left|L\right|}\right\} \right)\nonumber \\
 & \quad\times\sum_{p=1}^{\left|L\right|!}\prod_{j=1}^{\left|L\right|}\sum_{i=1}^{N_{L_{j}}^{-}}g_{L_{j}}^{i}\left(x_{\phi_{\left|L\right|,p,j}}\right)\label{eq:RFS_particles}
\end{align}
where $\left[\phi_{t,p,1},...,\phi_{t,p,t}\right]^{T}$ $p\in\left\{ 1,...,t!\right\} $
represents the permutations of vector $\left[1,...,t\right]^{T}$.
We recall from Theorem \ref{thm:Optimisation_RFS_family} that any
density $\varphi^{n}\left(\cdot;\overrightarrow{L}\right)$ of the
recursion can be written as $\varphi^{n}\left(\cdot;\overrightarrow{L}\right)=\alpha^{n}\left(;\overrightarrow{L}\right)\check{\pi}\left(\cdot;\overrightarrow{L}\right)$.
Using the auxiliary vector $a_{1:\left|L\right|}$, which was defined
in (\ref{eq:posterior_PP}), we get
\begin{align}
\varphi^{n}\left(x_{1:\left|L\right|},a_{1:\left|L\right|};\overrightarrow{L}\right) & \propto\alpha^{n}\left(x_{1:\left|L\right|};\overrightarrow{L}\right)\ell\left(\left\{ x_{1},...,x_{\left|L\right|}\right\} \right)\nonumber \\
 & \quad\times\sum_{p=1}^{\left|L\right|!}\prod_{j=1}^{\left|L\right|}g_{L_{j}}^{a_{j}}\left(x_{\phi_{\left|L\right|,p,j}}\right)\label{eq:fi_k_l_aux}
\end{align}
From the MC approximation $\left\{ \left(L^{1},x_{1:\left|L^{1}\right|}^{1},a_{1:\left|L^{1}\right|}^{1}\right),....,\left(L^{N},x_{1:\left|L^{N}\right|}^{N},a_{1:\left|L^{N}\right|}^{N}\right)\right\} $
to the density $\boldsymbol{\pi}$, which is obtained using the GPP
PF, we can directly get the PMF $P_{\pi}$ and rearrange the particles
to get the PDFs of the states and auxiliary variables given the label
set for the initial PDF:
\begin{align}
P_{\pi}\left(L\right) & =N_{L}/N\label{eq:PMF_existence_MCMC_ini}\\
\varphi^{0}\left(x_{1:\left|L\right|},a_{1:\left|L\right|};\overrightarrow{L}\right) & \propto\sum_{i=1}^{N_{L}}\delta\left(a_{1:\left|L\right|}-a_{1:\left|L\right|}^{L,i}\right)\nonumber \\
 & \quad\times\delta\left(x_{1:\left|L\right|}-x_{1:\left|L\right|}^{L,i,0}\right)\label{eq:PDF_MCMC_ini}
\end{align}
where $N_{L}$ is the number of particles with label set $L$ and
$\left(x_{1:\left|L\right|}^{L,i,n},a_{1:\left|L\right|}^{L,i}\right)$
denotes the $i$th particle of the PDF $\varphi^{n}\left(\cdot;\overrightarrow{L}\right)$.

The recursion can be performed if, given an MC approximations to $\varphi^{n}\left(\cdot;\overrightarrow{L}\right)$,
we can obtain an MC approximation to $\varphi^{n+1}\left(\cdot;\overrightarrow{L}\right)$
$\forall L$. In order to get samples from $\varphi^{n+1}\left(\cdot;\overrightarrow{L}\right)$,
see (\ref{eq:fi_k_l_aux}), we need to find $\alpha^{n+1}\left(\cdot\right)$
using Theorems \ref{thm:optimisation_independent} and \ref{thm:Optimisation_RFS_family}.
In Section \ref{sub:Approximation-of-alpha}, we explain how to approximate
$\alpha^{n+1}\left(\cdot\right)$. In Section \ref{sub:MCMC-steps-to-LMB},
we describe the MCMC algorithm to obtain samples from (\ref{eq:fi_k_l_aux})
once $\alpha^{n+1}\left(\cdot\right)$ is obtained.

\subsection{Approximation of \textmd{\normalsize{}$\alpha^{n+1}\left(\cdot\right)$\label{sub:Approximation-of-alpha}}}

In order to be able to run the MCMC algorithm that gives samples from
$\varphi^{n+1}\left(\cdot;\overrightarrow{L}\right)$, we need to
calculate $\alpha^{n+1}\left(\cdot\right)$ based on the MC approximation
to $\varphi^{n}\left(\cdot;\overrightarrow{L}\right)$ $\forall\overrightarrow{L}$
and Theorems \ref{thm:optimisation_independent} and \ref{thm:Optimisation_RFS_family}. 

Using (\ref{eq:PMF_existence_MCMC_ini}) and (\ref{eq:PDF_MCMC_ini})
and Theorem \ref{thm:optimisation_independent}, we get
\begin{align}
\nu^{n}\left(x_{1};j\right) & =\frac{1}{N_{j}}\sum_{L\ni j}\sum_{i=1}^{N_{L}}\delta\left(x_{1}-x_{j}^{L,i,n}\right)\label{eq:nu_pf_approx_individual}\\
N_{j} & =\sum_{L\ni j}N_{L}
\end{align}
where $x_{j}^{L,i,n}$ indicates the $i$th particle of the target
with label $j$ in $\varphi^{n}\left(\cdot;\overrightarrow{L}\right)$.
Theorem \ref{thm:Optimisation_RFS_family} provides $\alpha^{n+1}\left(\cdot\right)$
based on $\nu^{n}\left(\cdot;\overrightarrow{L}\right)=\prod_{j=1}^{\left|L\right|}\nu^{n}\left(\cdot;L_{j}\right)$.
However, we cannot evaluate $\nu^{n}\left(\cdot;\overrightarrow{L}\right)$
directly because the PDF is represented by Dirac delta functions.
We solve this by approximating (\ref{eq:nu_pf_approx_individual})
as a Gaussian PDF 
\begin{equation}
\nu^{n}\left(x_{1};j\right)\approx\mathcal{N}\left(x_{1};\overline{x}_{j}^{n},P_{j}^{n}\right)\label{eq:nu_k_l_Gauss}
\end{equation}
where $\overline{x}_{j}^{n}$ and $P_{j}^{n}$ are obtained by moment
matching
\begin{align}
\overline{x}_{j}^{n} & =\frac{1}{N_{j}}\sum_{L\ni j}\sum_{i=1}^{N_{L}}x_{j}^{L,i,n}\label{eq:mean_PF}\\
P_{j}^{n} & =\frac{1}{N_{j}}\sum_{L\ni j}\sum_{i=1}^{N_{L}}\left(x_{j}^{L,i,n}-\overline{x}_{j}^{n}\right)\left(x_{j}^{L,i,n}-\overline{x}_{j}^{n}\right)^{T}.\label{eq:cov_PF}
\end{align}
This approximation is accurate enough to improve the LMB approximation
in the examples of \cite{Angel14_b} and Section \ref{sec:Numerical-simulations}.
Otherwise, we can regularise (\ref{eq:nu_pf_approx_individual}) \cite{Ristic_book04}.
Using Theorem \ref{thm:Optimisation_RFS_family}, we get
\begin{align}
\alpha^{n+1}\left(x_{1:\left|L\right|};\overrightarrow{L}\right) & \approx\frac{\prod_{j=1}^{\left|L\right|}\nu^{n}\left(x_{j};L_{j}\right)}{\sum_{p=1}^{\left|L\right|!}\prod_{j=1}^{\left|L\right|}\nu^{n}\left(x_{\phi_{\left|L\right|,p,j}};L_{j}\right)}.\label{eq:alpha_mcmc_approx}
\end{align}

\subsection{MCMC steps to improve the LMB approximation\label{sub:MCMC-steps-to-LMB}}

In the previous subsection, we indicated how to approximate $\alpha^{n+1}\left(\cdot\right)$
based on the MC approximations to $\varphi^{n}\left(\cdot;\overrightarrow{L}\right)$.
In this section, we design an MCMC algorithm to sample from $\varphi^{n+1}\left(\cdot;\overrightarrow{L}\right)$,
which is given by (\ref{eq:fi_k_l_aux}). In principle, we can use
any MCMC algorithm, e.g., Metropolis-Hastings \cite{Liu_book01}.
Nevertheless, we can perform the MCMC sampling more efficiently due
to the characteristics of the PDF (\ref{eq:fi_k_l_aux}). 

In any MCMC algorithm, we evaluate the target PDF up to a proportionality
constant. When we evaluate (\ref{eq:fi_k_l_aux}) for a state $\left(x_{1:\left|L\right|},a_{1:\left|L\right|}\right)$,
we compute the same terms we would calculate to evaluate (\ref{eq:fi_k_l_aux})
for any of its permutations $\left(\Gamma_{p,\left|L\right|}\left(x_{1:\left|L\right|}\right),a_{1:\left|L\right|}\right)$
$p=1,...,\left|L\right|!$. Therefore, as evaluating (\ref{eq:fi_k_l_aux})
for all the permutations comes at practically no extra cost, it is
of high interest to develop an MCMC algorithm that accounts for all
the permutations at the same step. 

Conditioned on $a_{1:\left|L\right|}$, the proposed MCMC algorithm
performs moves in the state. The algorithm requires the specification
of a transition rule w.r.t. which the target PDF is invariant \cite{Liu_book01}.
For a given $x_{1:\left|L\right|}$, we propose the following transition
rule to the next state $y_{1:\left|L\right|}$. First, we sample $\tilde{x}_{1:\left|L\right|}$
from a density $q\left(\tilde{x}_{1:\left|L\right|}\left|x_{1:\left|L\right|}\right.\right)=\prod_{j=1}^{\left|L\right|}q'\left(\tilde{x}_{j}\left|x_{j}\right.\right)$.
Then, we set $y_{1:\left|L\right|}=\Gamma_{p,\left|L\right|}\left(x_{1:\left|L\right|}\right)$
or $y_{1:\left|L\right|}=\Gamma_{p,\left|L\right|}\left(\tilde{x}_{1:\left|L\right|}\right)$
with probabilities $\beta'_{p}$ and $\tilde{\beta}'_{p}$ for $p=1,...,t!$
\begin{equation}
y_{1:\left|L\right|}=\left\{ \begin{array}{cc}
\Gamma_{p,\left|L\right|}\left(x_{1:\left|L\right|}\right) & \beta'_{p}=\frac{\beta_{p}}{\sum_{p=1}^{t!}\left(\beta_{p}+\tilde{\beta}_{p}\right)}\\
\Gamma_{p,\left|L\right|}\left(\tilde{x}_{1:\left|L\right|}\right) & \tilde{\beta}'_{p}=\frac{\tilde{\beta}_{p}}{\sum_{p=1}^{t!}\left(\beta_{p}+\tilde{\beta}_{p}\right)}
\end{array}\right.\label{eq:Y_selection}
\end{equation}
where
\begin{align}
\beta_{p}= & \varphi^{n+1}\left(\Gamma_{p,\left|L\right|}\left(x_{1:\left|L\right|}\right),a_{1:\left|L\right|};\overrightarrow{L}\right)\label{eq:beta_p}\\
\tilde{\beta}_{p}= & \varphi^{n+1}\left(\Gamma_{p,\left|L\right|}\left(\tilde{x}_{1:\left|L\right|}\right),a_{1:\left|L\right|};\overrightarrow{L}\right)\label{eq:beta_tilde_p}
\end{align}
and $\varphi^{k,n+1}\left(\cdot\right)$ is given by (\ref{eq:fi_k_l_aux}).
It is shown in Appendix \ref{sec:AppendixD} that this transition
rule is invariant if $q'\left(\cdot\left|\cdot\right.\right)$ is
symmetric and therefore leads to a valid MCMC algorithm. 

To sum up, the ILMB-MCMC has three design parameters: the number $I_{1}$
of steps of the recursion explained in Section \ref{sec:Improvement-LMB},
the number $I_{2}$ of steps that allow for the MCMC burn-in period
to get samples from $\varphi^{k,n}\left(\cdot\right)$ and the transition
density $q\left(\cdot\left|x_{1:\left|L\right|}\right.\right)$. Its
steps are given in Algorithm \ref{alg:ILMB-MCMC_algorithm}.

\begin{algorithm}
\protect\caption{\label{alg:ILMB-MCMC_algorithm}MCMC algorithm to improve the LMB
approximation (ILMB-MCMC) }

{\fontsize{9}{9}\selectfont

\textbf{Input:} $N$ particles with even weights $\left\{ \left(L^{1},x_{1:\left|L^{1}\right|}^{1},a_{1:\left|L^{1}\right|}^{1}\right),....,\left(L^{N},x_{1:\left|L^{N}\right|}^{N},a_{1:\left|L^{N}\right|}^{N}\right)\right\} $,
which approximate (\ref{eq:posterior_PP}).

\textbf{Output:} Existence probability and $N_{j}$ particles which
represent the density of the target with label $j$.

\begin{algorithmic}     

\State - Obtain $P_{\pi}$ using (\ref{eq:PMF_existence_MCMC_ini}).

\ForAll{$L$ }

\State - Obtain $\left(x_{1:\left|L\right|}^{L,i,0},a_{1:\left|L\right|}^{L,i}\right)$,
see (\ref{eq:PDF_MCMC_ini}).

\EndFor

\For{ $n=0$ to $I_{1}-1$ }

\State - Calculate $\overline{x}_{j}^{n}$ and $P_{j}^{n}$ for $j\in\left\{ 1,...,\kappa\right\} $
using (\ref{eq:mean_PF}) and (\ref{eq:cov_PF}).

\ForAll{$L$ }

\State - Calculate $\alpha^{n+1}\left(\cdot;\overrightarrow{L}\right)$
using (\ref{eq:alpha_mcmc_approx}).

\State - Set $x_{1:\left|L\right|}^{L,i,n+1}=x_{1:\left|L\right|}^{L,i,n}\quad i=1,...,N_{L}$. 

\For{ $h=0$ to $I_{2}-1$}

\For{ $i=1$ to $N_{L}$}

\State - Sample $\tilde{x}_{1:\left|L\right|}^{i}$ from $q\left(\cdot\left|x_{1:\left|L\right|}^{L,i,n+1}\right.\right)$.

\State - Use $x_{1:\left|L\right|}^{L,i,n+1}$ and $\tilde{x}_{1:\left|L\right|}^{i}$
to calculate $\beta_{p}$ and $\tilde{\beta}_{p}$ by (\ref{eq:beta_p}),
(\ref{eq:beta_tilde_p}) and (\ref{eq:fi_k_l_aux}).

\State - Select $y_{1:\left|L\right|}$ according to (\ref{eq:Y_selection}). 

\State - Set $x_{1:\left|L\right|}^{L,i,n+1}=y_{1:\left|L\right|}$.

\EndFor

\EndFor

\EndFor

\EndFor

\State - Obtain the existence probabilities for $j=\left\{ 1,...,\kappa\right\} $
using $P_{\pi}$ and Theorem \ref{thm:optimisation_independent}.

\end{algorithmic}

}
\end{algorithm}

\subsection{Discussion\label{sub:Discussion}}

In this paper, we have developed a PF, which is called GPP, that approximates
the labelled posterior based on an LMB posterior at the previous time
step. This PF can be used on its own without the need of any MCMC
or label switching improvement algorithm. Nevertheless, if we are
interested in the unlabelled posterior, we can make use of the recursion
explained in Section \ref{sec:Improvement-LMB} to improve the LMB
approximation. Section \ref{sec:MCMC-algorithm} explains an algorithm
based on MCMC to approximate the recursion in Section \ref{sec:Improvement-LMB}.
The complexity of the MCMC algorithm depends on the number of targets
through the number of possible permutations. Nevertheless, the benefits
of the MCMC algorithm are only expected to happen when targets get
in close proximity and then separate because of the mixing of the
labels \cite{Boers10}. As a result, in practice, we should only apply
the MCMC algorithm for targets in close proximity, e.g., by clustering
\cite{Yi13,Morelande07,Angel13}.

We also want to mention that, in some cases, the dimensionality of
the PF can be reduced by applying Rao-Blackwellisation \cite{Morelande07}.
In this case, measurement non-linearities can only be a function of
some elements of the state, e.g., they apply to the position but not
the velocity. Function $\alpha^{n}\left(\cdot\right)$ in (\ref{eq:fi_k_l_aux})
would prevent the use of Rao-Blackwellisation even when the measurement
and dynamic models allow for it. This could be sorted out by looking
for a labelled PDF within the unlabelled RFS family that allows for
Rao-Blackwellisation.  Nevertheless, the development of this idea
is beyond the scope of this paper.

\section{Numerical simulations\label{sec:Numerical-simulations}}

In this section, we evaluate the performances of the GPP method and
the ILMB-MCMC algorithm in MTT using a sensor network. We analyse
two examples. In the first one, we set aside the problem of target
births and study the performance of the proposed algorithms when targets
get in close proximity and then separate. The second example considers
target births. 

Both examples are based on the following dynamic/measurement models.
In this section, we use a superindex to denote the time step $k$.
The state vector of the $j$th target at time $k$ is $x_{j}^{k}=\left[p_{x,j}^{k},\dot{p}_{x,j}^{k},p_{y,j}^{k},\dot{p}_{y,j}^{k}\right]^{T}$.
The dynamic model of the target is the nearly-constant velocity model:
\begin{align}
x_{j}^{k+1} & =Fx_{j}^{k}+v^{k}\\
F & =I_{2}\otimes\left(\begin{array}{cc}
1 & \tau\\
0 & 1
\end{array}\right)
\end{align}
where $\otimes$ is the Kronecker product, $I_{n}$ is the identity
matrix of size $n$ and $v^{k}$ is the process noise at time $k$.
The process noise is zero-mean Gaussian distributed with covariance
matrix
\begin{equation}
Q=qI_{2}\otimes\left(\begin{array}{cc}
\tau^{3}/3 & \tau^{2}/2\\
\tau^{2}/2 & \tau
\end{array}\right)
\end{equation}
where $q$ is a parameter of the model and $\tau$ is the sampling
period.

There are $M=252$ sensors, so the measurement vector at time $k$
is $z^{k}=\left[z_{1}^{k},...,z_{M}^{k}\right]^{T}$ where $z_{j}^{k}$
is the measurement of the $j$th sensor at time $k$. Sensor $m$
is located at $\left[\xi_{x,m},\xi_{y,m}\right]^{T}$ and measures
an acoustic signal emitted by the target with measurement model 
\begin{align}
z_{m}^{k}= & \sqrt{\sum_{j=1}^{t}h_{m}\left(x_{j}^{k}\right)}+w_{m}^{k}\label{eq:measurement_scenario}
\end{align}
where
\begin{align}
h_{m}\left(x_{j}^{k}\right)= & \left\{ \begin{array}{cc}
\frac{P_{0}d_{0}^{2}}{d_{m}^{2}\left(x_{j}^{k}\right)} & d_{m}^{2}\left(x_{j}^{k}\right)>d_{0}^{2}\\
P_{0} & d_{m}^{2}\left(x_{j}^{k}\right)\leq d_{0}^{2}
\end{array}\right.
\end{align}
and $w_{m}^{k}$ is an independent zero-mean Gaussian noise with variance
$\sigma_{s}^{2}$, $P_{0}$ is the saturation power, $d_{0}$ is the
distance at which this saturation power is produced and $d_{m}^{2}\left(x_{j}^{k}\right)$
is the square distance from the target $x_{j}^{k}$ to the $m$th
sensor
\begin{equation}
d_{m}^{2}\left(x_{j}^{k}\right)=\left(p_{x,j}^{k}-\xi_{x,m}\right)^{2}+\left(p_{y,j}^{k}-\xi_{y,m}\right)^{2}.
\end{equation}
Measurement equation (\ref{eq:measurement_scenario}) models the amplitude
of the received acoustic signal at a sensor from incoherently emitting
targets \cite{Sheng05} and is not superpositional so filters such
as \cite{Nannuru13,Papi15} cannot be used.

\subsection{Targets in close proximity for a long time\label{sub:Example-with-no-births}}

The main objective of this section is to analyse algorithm performances
when targets get in close proximity for a sufficiently long time and
then separate. In this case, there is mixing of the labels so the
labelling switching algorithm based on MCMC is expected to be useful.
In order to study this case in an isolated way, we assume that there
are no target births

We have implemented the GPP method followed by a usual MCMC algorithm
(U-MCMC GPP) after resampling and the GPP method with the ILMB-MCMC
algorithm. The U-MCMC algorithm corresponds to the Metropolis-Hastings
algorithm with 20 steps \cite{Ristic_book04}. We have implemented
3 versions of ILMB-MCMC GPP that differ in their parameters $I_{1}$
and $I_{2}$, see Table \ref{tab:IT-MCMC-parameters}. They have roughly
the same number of MCMC steps, indicated by $I_{1}\times I_{2}$,
but different number $I_{1}$ of iterations in the sequence of Section
\ref{sec:Improvement-LMB}. We recall that U-MCMC improves sample
diversity while ILMB-MCMC improves sample diversity and the required
LMB approximation. Therefore, we approximately use the same number
of MCMC steps in both methods such that the improvement is not due
to sample diversity but improvement in LMB approximation, which is
what we want to assess. U-MCMC and ILMB-MCMC use the transition density
\begin{equation}
q'\left(\tilde{x}_{j}^{k}\left|x_{j}^{k}\right.\right)=\mathcal{N}\left(\tilde{x}_{j}^{k};x_{j}^{k},Q\right).
\end{equation}
We compare these algorithms with the two-layer parallel partition
(PP) PF, which has been demonstrated to outperform a variety of filters
in track-before-detect applications \cite{Angel13}. For this filter,
we use the same parameters as in \cite{Angel13}. We have also implemented
the sampling importance resampling PF (SIR-PF) \cite{Arulampalam02}.

\begin{table}
\protect\caption{\label{tab:IT-MCMC-parameters}ILMB-MCMC GPP parameters}

\centering{}%
\begin{tabular}{c|cc}
\hline 
Version &
$I_{1}$ &
$I_{2}$\tabularnewline
\hline 
1 &
1 &
20\tabularnewline
2 &
2 &
10\tabularnewline
3 &
3 &
6\tabularnewline
\hline 
\end{tabular}
\end{table}

The estimation error is evaluated using the optimal subpattern assignment
(OSPA) metric with $c=120\,\mathrm{m}$, Euclidean distance and $p=2$
\cite{Schuhmacher08}. In order to estimate target states, we first
obtain the most likely cardinality from the PF. If the cardinality
is known and $c\rightarrow\infty$, the minimum MSOSPA (MMSOSPA) estimator
corresponds to the mean of the unlabelled RFS density in a Voronoi
region \cite{Guerriero10,Angel14}. Therefore, we integrate out the
labels of the particles and then use $k$-means clustering on the
particles as in \cite{Kreucher05} as an approximation to the MMSOSPA
estimator. Importantly, this clustering is only performed to estimate
target states, it does not affect the approximation of the posterior
PDF at the following time steps as done in \cite{Kreucher05}.

\begin{table}
\protect\caption{\label{tab:Parameters-sensor-network}Parameters of the simulation}

\centering{}%
\begin{tabular}{c|cccccc}
\hline 
Parameter &
$\tau$ &
$q$ &
$P_{0}$ &
$\sigma_{s}^{2}$ &
$d_{0}$ &
$\gamma$\tabularnewline
\hline 
Value &
0.5 s &
3.24 $\mathrm{m^{2}/s^{3}}$ &
15.85 W &
1 &
20 m &
0.99\tabularnewline
\hline 
\end{tabular}
\end{table}

\begin{figure}
\begin{centering}
\includegraphics[scale=0.5]{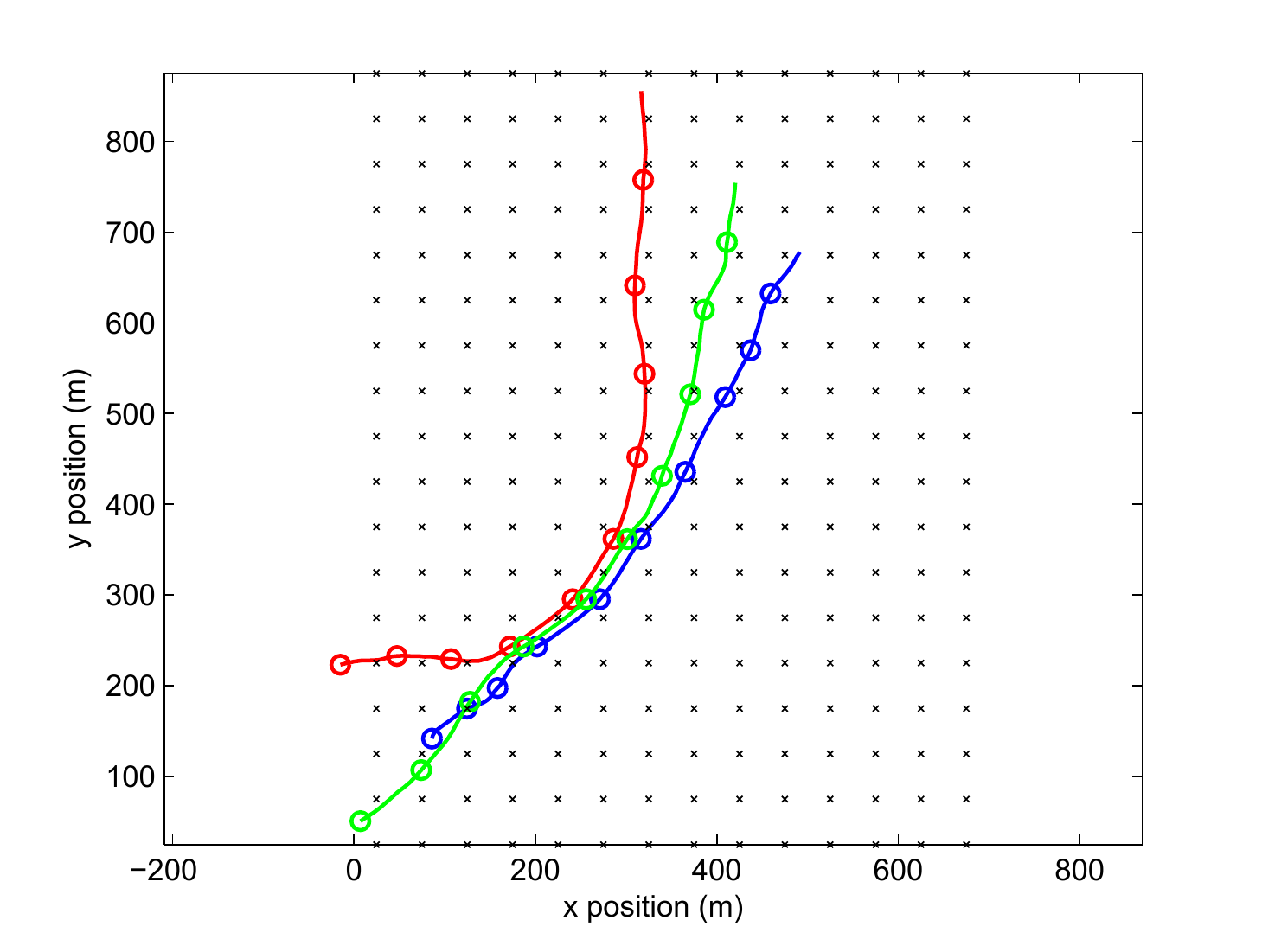}
\par\end{centering}

\protect\caption{\label{fig:Scenaro_no_birth}Scenario of the simulations. The target
positions every ten time steps is represented by a circumference.
The targets move from left to right. The sensor positions are represented
by black crosses. }
\end{figure}

The target trajectories are shown in Figure \ref{fig:Scenaro_no_birth}.
The target with blue trajectory in Figure \ref{fig:Scenaro_no_birth}
dies at time step 85 while the other two targets are alive at all
time steps. The algorithms' performances are evaluated based on Monte
Carlo simulation with 500 runs. The simulation parameters are those
given in Table \ref{tab:Parameters-sensor-network}, where we recall
that $\gamma$ is the probability of survival. The prior PDF of the
$j$th target at time step 0 is $\mathcal{N}\left(x_{j}^{0};\overline{x}_{j}^{0},\Sigma^{0}\right)$
where $\overline{x}_{j}^{0}$ and $\Sigma^{0}$ are the mean and covariance
matrix of the $j$th target at time 0. In each Monte Carlo run, $\overline{x}_{j}^{0}$
is drawn from Gaussian distribution whose mean is the true target
position and $\Sigma^{0}=100I_{4}$. In addition, the filters are
also initiated with another target, which does not exist, whose PDF
is Gaussian with mean $\overline{x}_{j}^{0}=\left[600,0,200,0\right]^{T}$
and covariance $100I_{4}$ in international system units. The initial
probability of existence of all targets is set to one.

\begin{figure}
\begin{centering}
\subfloat[]{\protect\begin{centering}
\protect\includegraphics[scale=0.5]{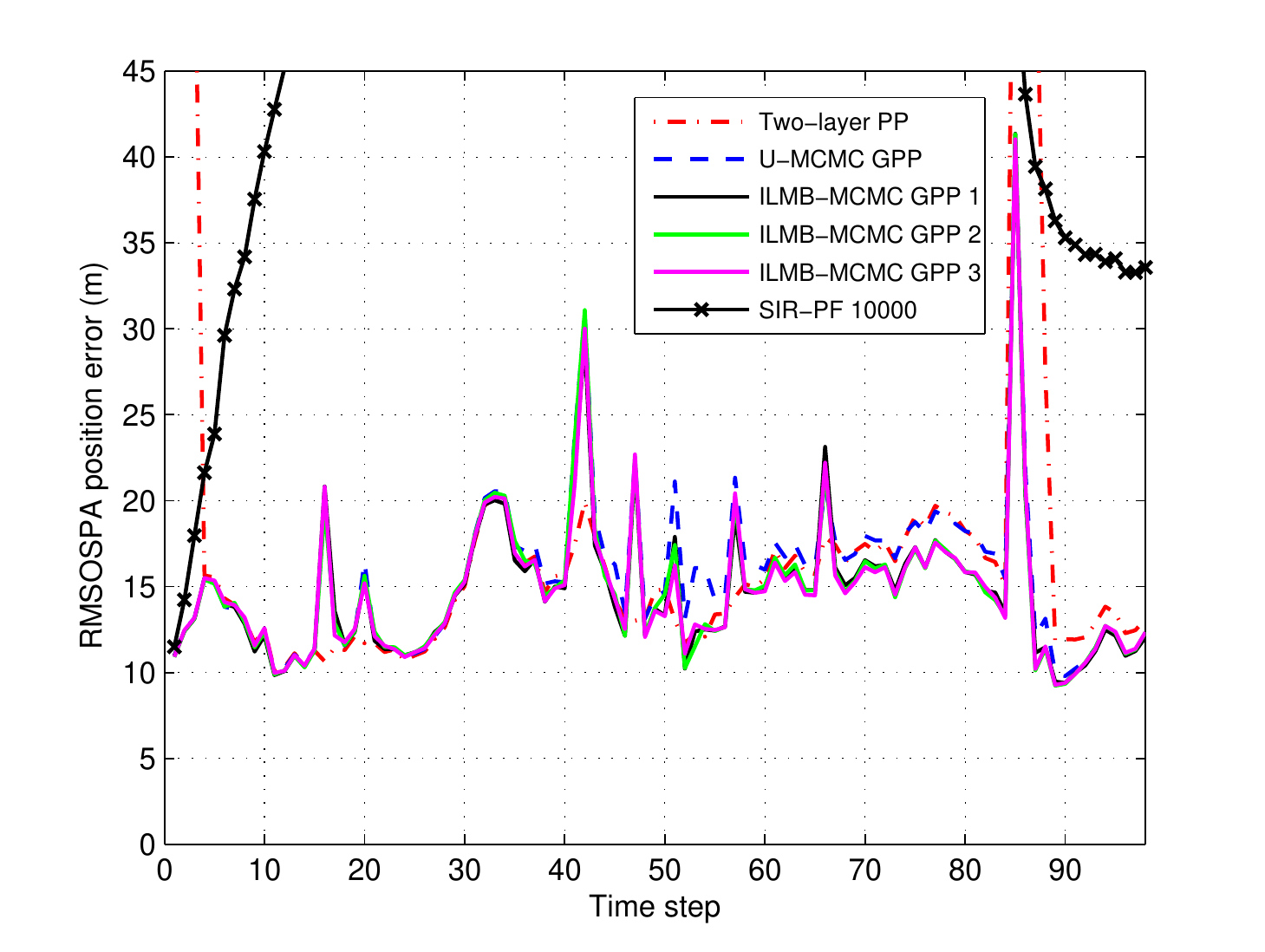}\protect
\par\end{centering}

}
\par\end{centering}

\begin{centering}
\subfloat[]{\protect\begin{centering}
\protect\includegraphics[scale=0.5]{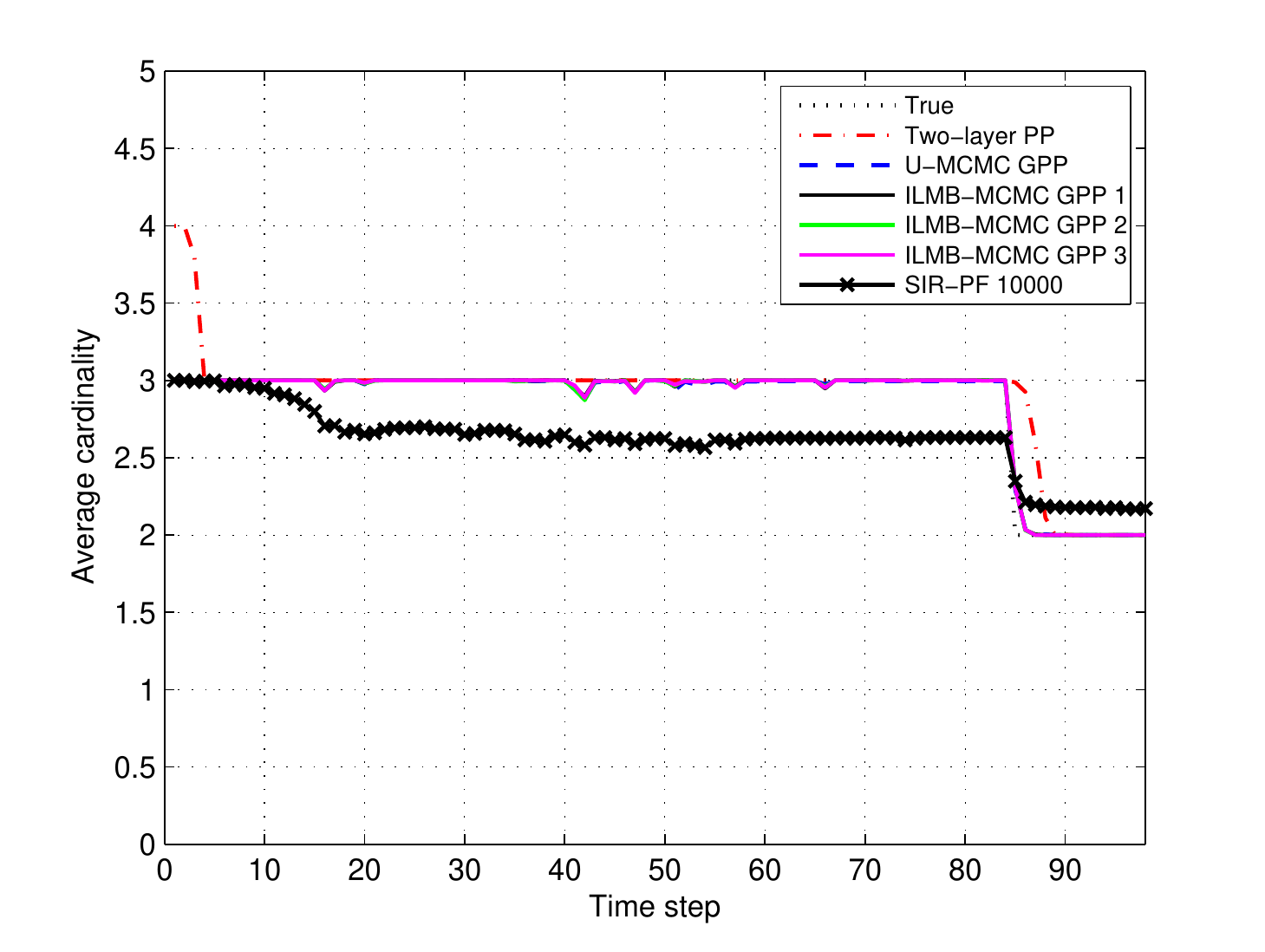}\protect
\par\end{centering}

}
\par\end{centering}

\protect\caption{\label{fig:Performance_scenario1}RMSOSPA position error (a) and average
cardinality (b) against time for the scenario in Fig. \ref{fig:Scenaro_no_birth}.
ILMB filters perform best especially after the targets have been in
close proximity for some time. The two-layer PP PF has an inherent
delay in removing tracks. SIR-PF with 10000 particles does not perform
well.}
\end{figure}

First, we assess the filters' performances using 500 particles except
for the SIR-PF, which uses 10000 particles. The aim of this is to
clearly demonstrate the benefits of using LMB assumption and how it
can improve filter performance. The root mean square OSPA (RMSOSPA)
position errors \cite{Rahmathullah16_prov} against time and average
cardinality are shown in Figure \ref{fig:Performance_scenario1}.
SIR-PF does not perform well even though it uses 10000 particles instead
of 500. The two-layer PF in \cite{Angel13} has a bias to remove targets
as it always requires several time steps. Therefore, error increases
significantly when a target disappears. ILMB-GPP methods do not have
this drawback and are able to provide a low error when targets disappear.
In Figure \ref{fig:Performance_scenario1} (a), there are several
spikes in the error for ILMB-GPP algorithms. These spikes arise due
to errors in cardinality estimation, as seen in Figure \ref{fig:Performance_scenario1}
(b). All the versions of GPP-ILMB provide an improvement over U-MCMC
specially after time step 40 until around time step 90. The former
corresponds to a time step in which the targets have been in close
proximity for a while, see Figure \ref{fig:Scenaro_no_birth}. The
GPP-ILMB implementations perform quite similarly. This implies that
in this scenario one step of the LMB improvement sequence, described
in Section \ref{sec:Improvement-LMB}, makes a difference but further
steps provide a negligible improvement. 

\begin{figure}
\begin{centering}
\includegraphics[scale=0.5]{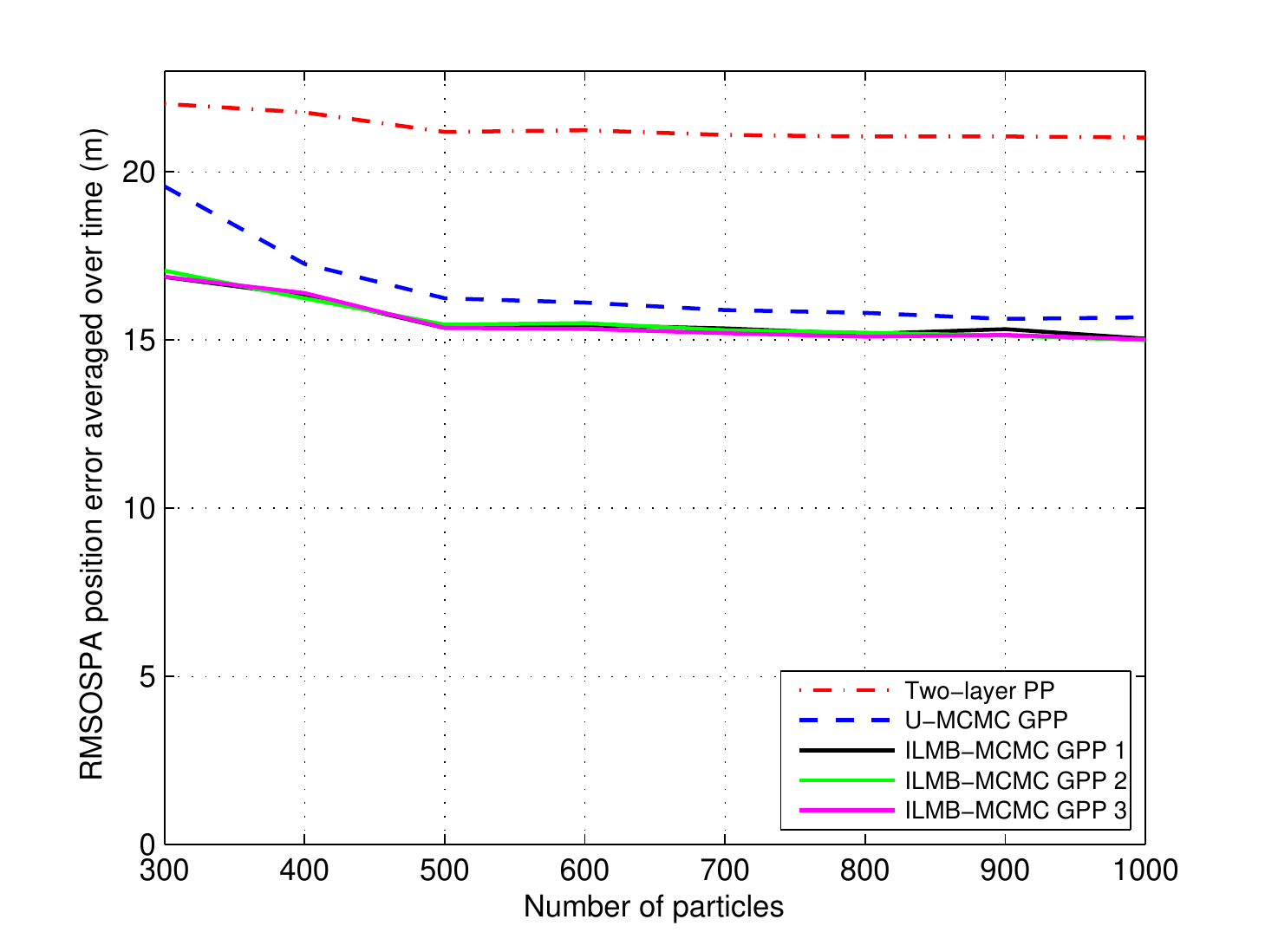}
\par\end{centering}

\protect\caption{\label{fig:Error_number_particles}RMSOSPA position error averaged
over time against the number of particles. ILMB performance is higher
than the rest of the algorithms for different number of particles.}
\end{figure}

Now, we analyse the effect of the number of particles on filter performance.
We show the RMSOSPA error averaged over time against the number of
particles in Figure \ref{fig:Error_number_particles}. Due to the
bias in target state estimation of the two-layer PF, increasing the
number of particles is not sufficient to lower the error. ILMB-MCMC
algorithms have similar performance and outperform the U-MCMC. As
expected, this improvement decreases as the particle number increases.
For example, for 300 particles, the difference in error between U-MCMC
and the best ILMB-MCMC is around 2.7 m while for 1000 particles it
is 0.6 m. In addition, we want to recall that ILMB-MCMC and U-MCMC
have basically the same error until time step 40 so the averaged error
over time does not really show the differences between the filter
performances when they really occur.

The computational complexity of ILMB-MCMC algorithms is slightly higher
than for U-MCMC and the two-layer PF. For example, for 500 particles,
the execution times in seconds of our Matlab implementation of the
algorithms on an Intel Core i7 laptop are: Two-layer PP (11.4), U-MCMC
GPP (10.29), ILMB-MCMC GPP 1 (12.1), 2 (12.7), 3 (12.0). Nevertheless,
according to Figure \ref{fig:Error_number_particles}, the error achieved
by ILMB method using 500 particles is not achieved by U-MCMC even
with 1000 particles. This implies that for a given objective error,
we can lower the number of particles and the execution times if we
use the MCMC algorithm developed in this paper. We also want to remark
that the execution time of the SIR-PF with 10000 particles is 12.5s
and has a significantly worse performance than the previous filters.

\subsection{Example with target births\label{sub:Example-with-target-births}}

We proceed to analyse the performance of the proposed algorithms in
the scenario shown in Figure \ref{fig:Scenario-target-births}, which
has target births, deaths and crossings though no targets are in close
proximity for a long time. Therefore, as was demonstrated in the previous
simulations, the improvement of the MCMC algorithm is expected to
be negligible as there is no mixing of labels \cite{Boers10}. The
objectives of this example are: show that the MCMC algorithm is only
needed if there is mixing of labels and assess how the filters work
if there are target births.

We use the same parameters as in the previous section. There are four
possible locations of target births with birth probability $10^{-3}$
and densities $\eta\left(\cdot;j\right)=\mathcal{N}\left(\cdot;,m_{j},100I_{4}\right)$,
with $m_{1}=\left(85,2,140,2\right)$, $m_{2}=\left(250,0,280,0\right)$,
$m_{3}=\left(145,0,575,0\right)$ and $m_{4}=\left(420,0,200,0\right)$
in international system units.

The proposed algorithms are implemented with 500 particles while the
SIR-PF uses 10000 particles. We do not compare with the two-layer
particle filter in \cite{Angel13} as it was designed for uniform
birth models not for this kind of birth model. GPP has been implemented
with and without U-MCMC and in this scenario both have similar performance.
The RMSOSPA position error and average cardinality for the algorithms
are shown in Figure \ref{fig:Performance_target_births}. As before,
SIR-PF has a much lower performance than the proposed algorithms.
GPP and ILMB-GPP algorithms perform similarly. As indicated before,
this is expected as targets are not in close proximity for a long
time, as in Figure \ref{fig:Scenaro_no_birth}, so the effect of ILMB-MCMC
is expected to be negligible as there is no mixing of labels \cite{Boers10}.
GPP methods estimate the cardinality quite accurately at most time
steps and provide low position errors. 

In this case the execution times in seconds are: GPP (4.8), ILMB-MCMC
GPP 1 (25.2), 2 (25.3), 3 (25.3) and SIR-PF (40). SIR-PF has a higher
computational complexity because of the number of particles and possible
labels it generates. The GPP method is able to sample labels more
efficiently by the importance density (\ref{eq:q_existence}). ILMB-MCMC
algorithms have a high computational burden w.r.t. GPP because we
perform the MCMC algorithm on all possible targets, which is inefficient.
As we mention in Section \ref{sub:Discussion}, in practice, we would
apply clustering so that we only perform the MCMC steps for the targets
that exhibit label mixing. 

\begin{center}
\begin{figure}
\begin{centering}
\includegraphics[scale=0.5]{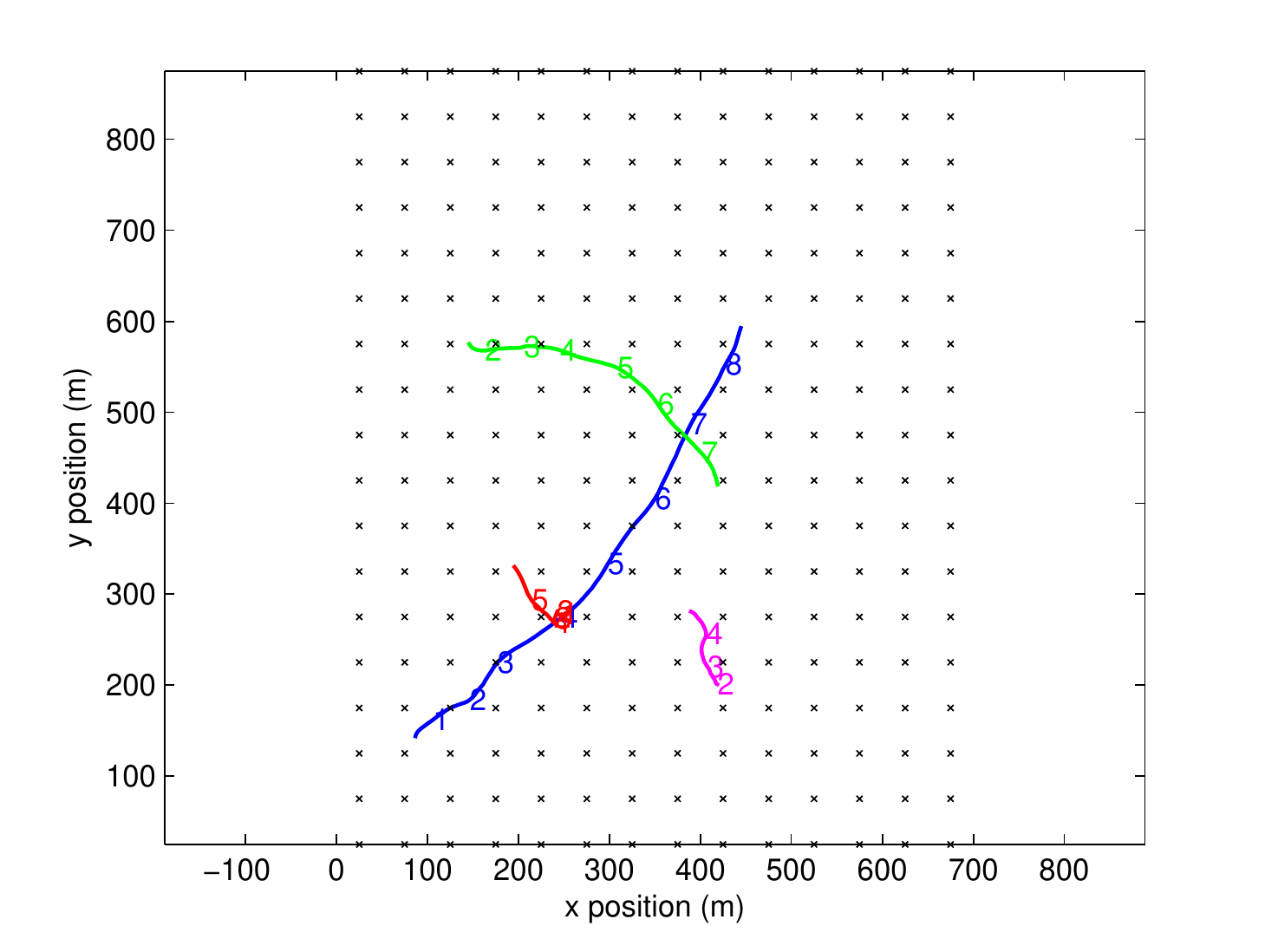}
\par\end{centering}

\protect\caption{\label{fig:Scenario-target-births}Scenario of the simulations with
target births. Target positions at time step $10k$ are marked by
$k$. Targets are born at time steps $\left(5,10,15,20\right)$ and
they die at time steps $\left(90,60,80,50\right)$. The sensor positions
are represented by black crosses. }

\end{figure}

\par\end{center}

\begin{figure}
\begin{centering}
\subfloat[]{\protect\begin{centering}
\protect\includegraphics[scale=0.5]{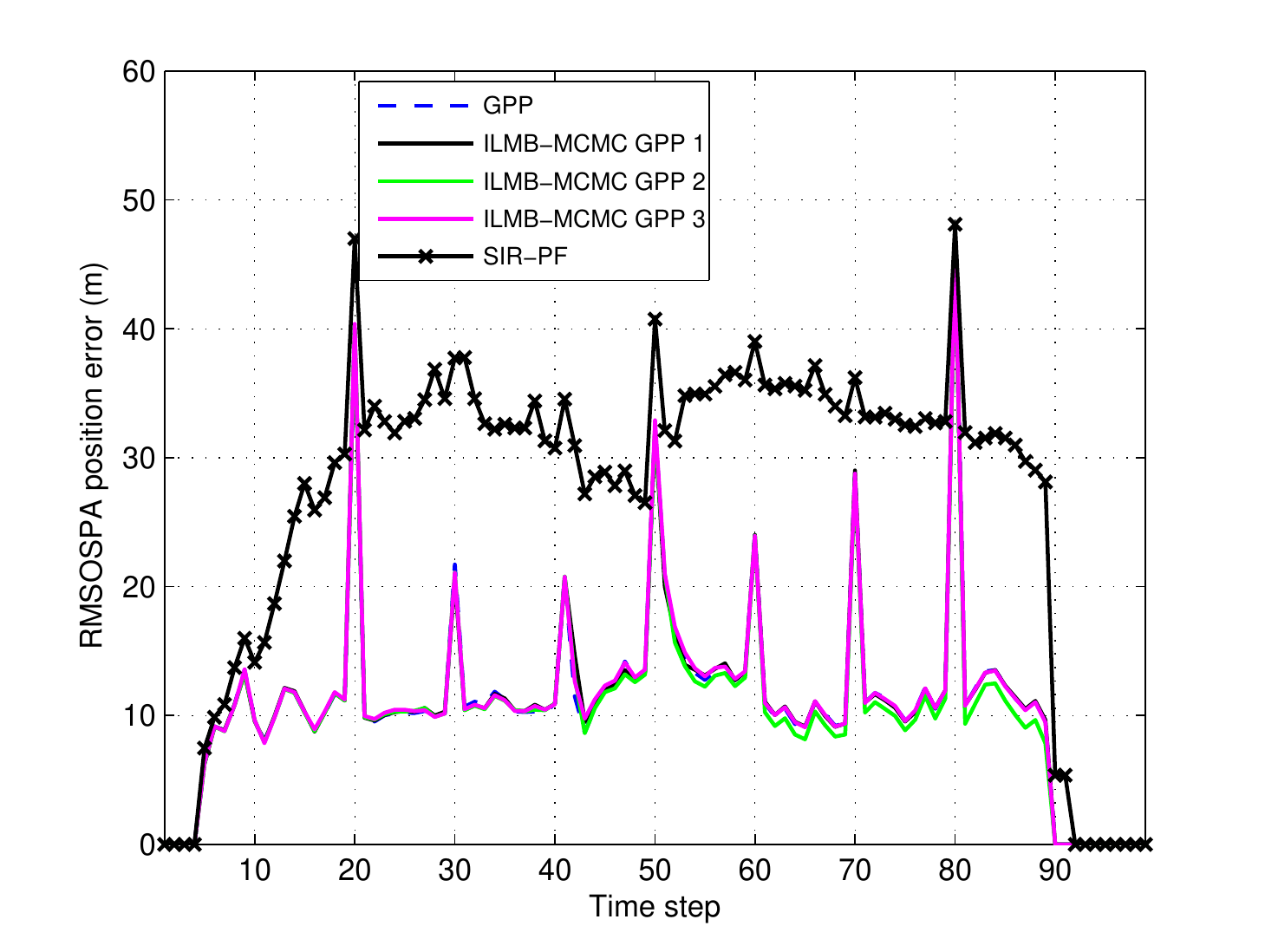}\protect
\par\end{centering}

}
\par\end{centering}

\begin{centering}
\subfloat[]{\protect\begin{centering}
\protect\includegraphics[scale=0.5]{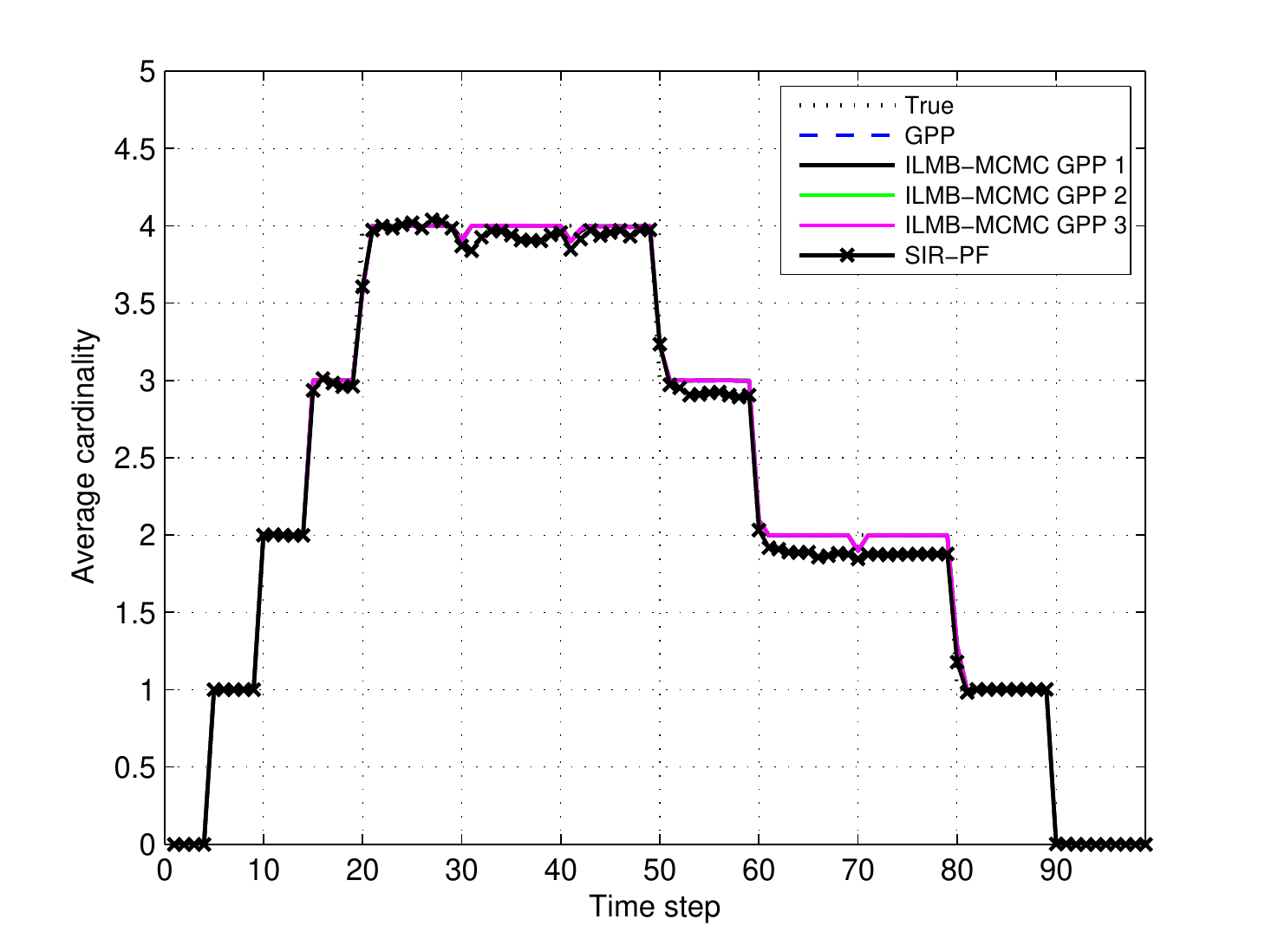}\protect
\par\end{centering}

}
\par\end{centering}

\protect\caption{\label{fig:Performance_target_births}RMSOSPA position error (a) and
average cardinality (b) against time for the scenario in Fig. \ref{fig:Scenario-target-births}.
GPP methods with 500 particles outperform SIR-PF with 10000 particles. }
\end{figure}

\section{Conclusions\label{sec:Conclusions}}

We have proposed the GPP-PF, which is a PF for general track-before-detect
models under an LMB assumption. The GPP-PF can be applied on its own
without the need of a label switching improvement algorithm, has a
low computational burden and efficiently deals with target births
and deaths.

We have also derived a label switching improvement algorithm. This
kind of algorithm can be applied when we use a labelled posterior
to approximate the unlabelled posterior. It is based on a sequence
of PDFs that improves the LMB approximation from a member of the unlabelled
RFS family of the posterior based on recursive KLD optimisations.
The label switching improvement algorihtm has been implemented and
incorporated to the GPP-PF using an MCMC algorithm called ILMB-MCMC.
The benefits of the ILMB-MCMC algorithm are expected to happen when
targets get in close proximity for a long time and separate. As a
result, if targets are not in close proximity for a sufficiently long
time, we should use the GPP-PF without ILMB-MCMC to save computational
resources. If targets get in close proximity for a sufficiently long
time, we should use the GPP-PF along with ILMB-MCMC to improve tracking
performance.

Future work should address how to perform clustering efficiently such
that we only perform the ILMB-MCMC steps for the targets with labelling
mixing. In addition, the algorithms should be adapted for Rao-Blackwellisation.

\appendices{}

\section{\label{sec:AppendixA}}

In this appendix, we prove (\ref{eq:KLD}). We have that
\begin{align}
\mathrm{D}\left(\boldsymbol{\pi}\left\Vert \boldsymbol{\nu}\right.\right) & =\int\boldsymbol{\pi}\left(\mathbf{X}\right)\log\frac{\boldsymbol{\pi}\left(\mathbf{X}\right)}{\boldsymbol{\nu}\left(\mathbf{X}\right)}\delta\mathbf{X}\nonumber \\
 & =\sum_{t=0}^{\infty}\frac{1}{t!}\sum_{\left[l_{1},...,l_{t}\right]\in\mathbb{L}^{t}}\int\boldsymbol{\pi}\left(\left\{ \left(x_{1},l_{1}\right),...,\left(x_{t},l_{t}\right)\right\} \right)\nonumber \\
 & \quad\times\log\frac{\boldsymbol{\pi}\left(\left\{ \left(x_{1},l_{1}\right),...,\left(x_{t},l_{t}\right)\right\} \right)}{\boldsymbol{\nu}\left(\left\{ \left(x_{1},l_{1}\right),...,\left(x_{t},l_{t}\right)\right\} \right)}dx_{1:t}.
\end{align}

Using (\ref{eq:pdf_conditioned_states}), we get
\begin{align}
 & \mathrm{D}\left(\boldsymbol{\pi}\left\Vert \boldsymbol{\nu}\right.\right)\nonumber \\
 & \quad=\sum_{t=0}^{\infty}\frac{1}{t!}\sum_{l_{1:t}\in\mathbb{L}^{t}}\int\pi\left(x_{1:t};l_{1:t}\right)P_{\pi}\left(\left\{ l_{1},...,l_{t}\right\} \right)\nonumber \\
 & \quad\quad\times\log\frac{\pi\left(x_{1:t};l_{1:t}\right)P_{\pi}\left(\left\{ l_{1},...,l_{t}\right\} \right)}{\nu\left(x_{1:t};l_{1:t}\right)P_{\nu}\left(\left\{ l_{1},...,l_{t}\right\} \right)}dx_{1:t}\nonumber \\
 & \quad=\sum_{t=0}^{\infty}\frac{1}{t!}\sum_{l_{1:t}\in\mathbb{L}^{t}}\left[P_{\pi}\left(\left\{ l_{1},...,l_{t}\right\} \right)\log\frac{P_{\pi}\left(\left\{ l_{1},...,l_{t}\right\} \right)}{P_{\nu}\left(\left\{ l_{1},...,l_{t}\right\} \right)}\right.\nonumber \\
 & \quad\quad\left.+P_{\pi}\left(\left\{ l_{1},...,l_{t}\right\} \right)\int\pi\left(x_{1:t};l_{1:t}\right)\log\frac{\pi\left(x_{1:t};l_{1:t}\right)}{\nu\left(x_{1:t};l_{1:t}\right)}dx_{1:t}\right]\nonumber \\
 & \quad=\mathrm{D}\left(P_{\pi}\left\Vert P_{\nu}\right.\right)+\sum_{t=0}^{\infty}\frac{1}{t!}\sum_{l_{1:t}\in\mathbb{L}^{t}}P_{\pi}\left(\left\{ l_{1},...,l_{t}\right\} \right)\nonumber \\
 & \quad\quad\times\mathrm{D}\left(\pi\left(\cdot;l_{1:t}\right)\left\Vert \nu\left(\cdot;l_{1:t}\right)\right.\right).
\end{align}
As the last term is permutation invariant w.r.t. $l_{1:t}$, we get
(\ref{eq:KLD}).

\section{\label{sec:AppendixB}}

In this appendix we prove Theorem \ref{thm:optimisation_independent}.
Due to the fact that the constraints are not coupled, we can first
obtain the best PMF of the labels and then the PDF of the states given
the labels. Then, we consider
\begin{align}
\underset{P_{\nu}}{\arg\min}\, & \mathrm{D}\left(P_{\varphi}\left\Vert P_{\nu}\right.\right)\label{eq:minimisation_label_PMF_append}
\end{align}
subject to (\ref{eq:constraint_ind_existences}). Substituting (\ref{eq:constraint_ind_existences})
into (\ref{eq:minimisation_label_PMF_append}), we can rewrite the
optimisation problem in terms of the parameters $p_{1},...,p_{\kappa}$
subject to the constraint $0\leq p_{i}\leq1$ \foreignlanguage{british}{$i\in\left\{ 1,...,\kappa\right\} $}.
According to (\ref{eq:constraint_ind_existences}), for the sets in
which $i$ is included, $P_{\nu}$ is proportional to $p_{i}$ and
otherwise it is proportional to $1-p_{i}$, then, we have to minimise
\begin{align*}
\underset{p_{1},...,p_{\kappa}}{\arg\min}\,\sum_{i=1}^{\kappa}f\left(p_{i}\right)
\end{align*}
where
\begin{align}
f\left(p_{i}\right) & =\mathrm{ct}-\sum_{L\subseteq\mathbb{L}:i\in L}\log\left(p_{i}\right)P_{\varphi}\left(L\right)\nonumber \\
 & \quad+\sum_{L\subseteq\mathbb{L}:i\notin L}\log\left(1-p_{i}\right)P_{\varphi}\left(L\right)
\end{align}
for all $i\in\mathbb{L}$ subject to $0\leq p_{i}\leq1$, where $\mathrm{ct}$
represents all the terms that do not depend on $p_{i}$. Function
$f\left(\cdot\right)$ is convex as it has the form of a KLD over
a PMF of a discrete variable with two states \cite{Boyd_book04}.
Calculating the derivative and setting it to zero, we get
\begin{align}
p_{i} & =\sum_{L\subseteq\mathbb{L}:i\in L}P_{\varphi}\left(L\right)
\end{align}
which completes the proof as $0\leq p_{i}\leq1$. 

Second, for a given collection of PDFs $\varphi\left(x_{1:t};l_{1:t}\right)$,
$\left\{ l_{1},...,l_{t}\right\} \subseteq\left\{ 1,...,\kappa\right\} $,
we obtain $\nu\left(\cdot;1\right)...\nu\left(\cdot;\kappa\right)$,
which meet (\ref{eq:constraint_ind_PDFs}), that minimise
\begin{align*}
 & \sum_{L\subseteq\mathbb{L}}P_{\varphi}\left(L\right)\int\varphi\left(x_{1:\left|L\right|};\overrightarrow{L}\right)\log\frac{\varphi\left(x_{1:\left|L\right|};\overrightarrow{L}\right)}{\nu\left(x_{1:\left|L\right|};\overrightarrow{L}\right)}dx_{1:\left|L\right|}
\end{align*}
We obtain the minimisation for $\nu\left(\cdot;1\right)$ although
the result is general. Expanding the logarithm, we want to minimise
the functional
\begin{align}
 & A\left[\nu\left(\cdot;1\right)\right]\nonumber \\
 & \quad=\mathrm{ct}-\sum_{L\subseteq\mathbb{L}:1\in L}P_{\varphi}\left(L\right)\int\varphi_{1}\left(x_{1};\overrightarrow{L}\right)\log\nu\left(x_{1};1\right)dx_{1}
\end{align}
where $\mathrm{ct}$ represents all the terms that do not depend on
$\nu\left(\cdot;1\right)$ and $\varphi_{1}\left(\cdot;\overrightarrow{L}\right)$
is the marginal PDF of the first variable of $\varphi\left(\cdot;\overrightarrow{L}\right)$.
By KLD minimisation over vector spaces, this functional is minimised
by \cite{Bishop_book06} 
\begin{align}
\nu\left(x_{1};1\right) & \propto\sum_{L\subseteq\mathbb{L}:1\in L}P_{\varphi}\left(L\right)\varphi_{1}\left(x_{1};\overrightarrow{L}\right).
\end{align}

\section{\label{sec:AppendixC}}

In this appendix, we prove Theorem \ref{thm:Optimisation_RFS_family}.
Given $\boldsymbol{\varphi}$, its corresponding unlabelled density
is \cite{Vo13} 
\begin{align}
 & \check{\varphi}\left(\left\{ x_{1},...,x_{t}\right\} \right)\nonumber \\
 & \quad=\sum_{l_{1:t}\in\mathbb{L}^{t}}\boldsymbol{\varphi}\left(\left\{ \left(x_{1},l_{1}\right),...,\left(x_{t},l_{t}\right)\right\} \right)\nonumber \\
 & \quad=\sum_{l_{1:t}\in\mathbb{L}^{t}}P_{\varphi}\left(\left\{ l_{1},...,l_{t}\right\} \right)\varphi\left(x_{1:t};l_{1:t}\right)\nonumber \\
 & \quad=\sum_{\begin{subarray}{c}
\begin{subarray}{c}
\left\{ l_{1},...,l_{t}\right\} \subseteq\mathbb{L}\\
l_{1}<...<l_{t}
\end{subarray}\end{subarray}}\sum_{p=1}^{t!}P_{\varphi}\left(\left\{ \Gamma_{p,t}\left(l_{1},...,l_{t}\right)\right\} \right)\varphi\left(x_{1:t};\Gamma_{p,t}\left(l_{1:t}\right)\right)\nonumber \\
 & \quad=\sum_{\begin{subarray}{c}
\begin{subarray}{c}
\left\{ l_{1},...,l_{t}\right\} \subseteq\mathbb{L}\\
l_{1}<...<l_{t}
\end{subarray}\end{subarray}}P_{\varphi}\left(\left\{ l_{1},...,l_{t}\right\} \right)\check{\varphi}\left(\left\{ x_{1},...x_{t}\right\} ;l_{1:t}\right).
\end{align}

Given $\check{\boldsymbol{\nu}}$, $P_{\varphi}=P_{\pi}$, $L$ and
$\varphi\left(\cdot;\overrightarrow{L'}\right)$ $\forall L'\neq L$,
we want to obtain $\varphi\left(\cdot;\overrightarrow{L}\right)$
that minimises $\mathrm{D}\left(\check{\boldsymbol{\varphi}}\left\Vert \check{\boldsymbol{\nu}}\right.\right)$
subject to $\boldsymbol{\varphi}\in\left[\boldsymbol{\pi}\right]$.
Due to the way the sequence is constructed, constraint $\boldsymbol{\varphi}\in\left[\boldsymbol{\pi}\right]$
can be written as
\begin{equation}
\check{\varphi}\left(\left\{ x_{1},...,x_{t}\right\} ;\overrightarrow{L}\right)=\check{\pi}\left(\left\{ x_{1},...,x_{t}\right\} ;\overrightarrow{L}\right).\label{eq:constraint_set_append}
\end{equation}

In the KLD, there is only one term that depends on $\varphi\left(\cdot;\overrightarrow{L}\right)$
so we only need to minimise the functional
\begin{align}
A\left[\varphi\left(\cdot;\overrightarrow{L}\right)\right]= & \int\varphi\left(x_{1:t};\overrightarrow{L}\right)\log\frac{\varphi\left(x_{1:t};\overrightarrow{L}\right)}{\nu\left(x_{1:t};\overrightarrow{L}\right)}dx_{1:t}
\end{align}
subject to constraint (\ref{eq:constraint_set_append}). This minimisation
was solved in \cite[Appendix C]{Angel14_b} and its result is provided
in Theorem \ref{thm:Optimisation_RFS_family}. A quite similar proof
for a two-target case with Gaussian PDFs is found in \cite{Svensson10}.

\section{\label{sec:AppendixD}}

In this appendix we prove that the target PDF is invariant w.r.t.
the transition rule of the MCMC algorithm in Section \ref{sub:MCMC-steps-to-LMB}.
Here, we denote the transition PDF $\pi\left(\cdot\right)$ and the
transition density to go from $x_{1:t}$ to $y_{1:t}$ as $A\left(x_{1:t},y_{1:t}\right)$
and the number of targets is $t$. This is proved by the detailed
balance condition \cite{Liu_book01} 
\begin{align}
\pi\left(x_{1:t}\right)A\left(x_{1:t},y_{1:t}\right)= & \pi\left(y_{1:t}\right)A\left(y_{1:t},x_{1:t}\right)\label{eq:detailed_balance_append}
\end{align}
where $\pi\left(\cdot\right)$ denotes a general objective PDF, e.g.,
in Section \ref{sub:MCMC-steps-to-LMB}, this corresponds to $\varphi^{n}\left(x_{1:\left|L\right|}\left|a_{1:\left|L\right|}\right.;\overrightarrow{L}\right)$
as the MCMC moves are only proposed in variable $x_{1:\left|L\right|}$. 

According to Section \ref{sub:MCMC-steps-to-LMB}, we can write the
transition density $A\left(x_{1:t},y_{1:t},\right)$ as
\begin{align}
 & A\left(x_{1:t},y_{1:t},\right)\nonumber \\
 & \quad=\sum_{p=1}^{t!}\frac{q\left(\Gamma_{p,t}\left(y_{1:t}\right)\left|x_{1:t}\right.\right)}{t!}r\left(x_{1:t},y_{1:t}\right)\nonumber \\
 & \quad+\left(1-\int\sum_{p=1}^{t!}\frac{q\left(\Gamma_{p,t}\left(y_{1:t}\right)\left|x_{1:t}\right.\right)}{t!}r\left(x_{1:t},y_{1:t}\right)dy_{1:t}\right)\nonumber \\
 & \quad\times\frac{1}{\sum_{p=1}^{t!}\pi\left(\Gamma_{p,t}\left(x_{1:t}\right)\right)}\sum_{p=1}^{t!}\pi\left(y_{1:t}\right)\delta\left(y_{1:t}-\Gamma_{p,t}\left(x_{1:t}\right)\right)
\end{align}
where the acceptance probability of $y_{1:t}\neq\Gamma_{p,t}\left(x_{1:t}\right)$
is
\begin{align}
r\left(x_{1:t},y_{1:t}\right) & =\frac{\pi\left(y_{1:t}\right)}{\sum_{p=1}^{t!}\left[\pi\left(\Gamma_{p,t}\left(y_{1:t}\right)\right)+\pi\left(\Gamma_{p,t}\left(x_{1:t}\right)\right)\right]}.
\end{align}

Then, 
\begin{align}
 & \pi\left(x_{1:t}\right)A\left(x_{1:t},y_{1:t}\right)\nonumber \\
 & \quad=\sum_{p=1}^{t!}\frac{q\left(\Gamma_{p,t}\left(y_{1:t}\right)\left|x_{1:t}\right.\right)}{t!}\nonumber \\
 & \quad\quad\times\frac{\pi\left(y_{1:t}\right)\pi\left(x_{1:t}\right)}{\sum_{p=1}^{t!}\left[\pi\left(\Gamma_{p,t}\left(y_{1:t}\right)\right)+\pi\left(\Gamma_{p,t}\left(x_{1:t}\right)\right)\right]}\nonumber \\
 & \quad\quad+\left(1-\int\sum_{p=1}^{t!}\frac{q\left(\Gamma_{p,t}\left(y_{1:t}\right)\left|x_{1:t}\right.\right)}{t!}r\left(x_{1:t},y_{1:t}\right)dy_{1:t}\right)\nonumber \\
 & \quad\quad\times\frac{\pi\left(y_{1:t}\right)\pi\left(x_{1:t}\right)}{\sum_{p=1}^{t!}\pi\left(\Gamma_{p,t}\left(x_{1:t}\right)\right)}\sum_{p=1}^{t!}\delta\left(y_{1:t}-\Gamma_{p,t}\left(x_{1:t}\right)\right).\label{eq:append_d1}
\end{align}
As $q\left(y_{1:t}\left|x_{1:t}\right.\right)=\prod_{j=1}^{t!}q'\left(y_{j}\left|x_{j}\right.\right)$,
if $q'\left(\cdot\left|\cdot\right.\right)$ is symmetric, we get
(\ref{eq:detailed_balance_append}), which completes the proof. 

\bibliographystyle{IEEEtran}
\bibliography{13E__Trabajo_Angel_Mis_articulos_Improved_LMB_approx_Accepted_Referencias}

\end{document}